\documentclass[twocolumn,aps,prb,showpacs,showkeys,superscriptaddress,citeautoscript]{revtex4}
\usepackage[latin9]{inputenc}
\usepackage{color}
\usepackage{bm}
\usepackage{amsmath}
\usepackage{amssymb}
\usepackage{graphicx}

\makeatletter
\@ifundefined{textcolor}{}
{%
 \definecolor{BLACK}{gray}{0}
 \definecolor{WHITE}{gray}{1}
 \definecolor{RED}{rgb}{1,0,0}
 \definecolor{GREEN}{rgb}{0,1,0}
 \definecolor{BLUE}{rgb}{0,0,1}
 \definecolor{CYAN}{cmyk}{1,0,0,0}
 \definecolor{MAGENTA}{cmyk}{0,1,0,0}
 \definecolor{YELLOW}{cmyk}{0,0,1,0}
}

\usepackage{graphics}\usepackage{color}\usepackage{dcolumn}
\usepackage{bm}

\usepackage{amsfonts}
\usepackage{dcolumn}
\usepackage{bm}
\usepackage{color}
\usepackage{float}

\def\2{\frac{1}{2}} \def\4{\frac{1}{4}}
\def\6{\partial}
\def\+{\dagger}
\def\<{\langle} \def\>{\rangle}

\DeclareMathOperator{\sh}{sh}
\renewcommand{\sinh}{\sh}
\DeclareMathOperator{\ch}{ch}
\renewcommand{\cosh}{\ch}

\makeatother

\begin{document}

\title{On the absence of actual plateaus in zero-temperature magnetization
curves of quantum spin clusters and chains}

\author{Vadim Ohanyan}

\affiliation{Department of Theoretical Physics, Yerevan State University, Alex
Manoogian 1, 0025 Yerevan, Armenia}

\affiliation{ICTP, Strada Costiera 11, I-34151 Trieste, Italy}

\author{Onofre Rojas}

\affiliation{ICTP, Strada Costiera 11, I-34151 Trieste, Italy}

\affiliation{Departamento de Fisica, Universidade Federal de Lavras, CP 3037,
37200000, Lavras, MG, Brazil}

\author{Jozef Stre\v{c}ka}

\affiliation{Department of Theoretical Physics and Astrophysics, Faculty of Science,
P. J. Šafárik University, Park Angelinum 9, 040 01, Košice, Slovak
Republic}

\author{Stefano Bellucci}

\affiliation{INFN-Laboratori Nazionali di Frascati, Via E. Fermi 40, 00044 Frascati,
Italy}
\begin{abstract}
We examine the general features of the non-commutativity of the magnetization
operator and Hamiltonian for the small quantum spin clusters. The
source of this non-commutativity can be a difference in the Landé
g-factors for different spins in the cluster, XY-anisotropy in the
exchange interaction and the presence of the Dzyaloshinskii-Moriya
term in the direction different from the direction of the magnetic
field. As a result,  zero-temperature magnetization curves for small
spin clusters mimic those for the macroscopic systems with the band(s)
of magnetic excitations, i.e. for the given eigenstate of the spin
cluster the corresponding magnetic moment can be an explicit function
of the external magnetic field yielding the non-constant (non-plateau)
form of the magnetization curve within the given eigenstate. In addition,
the XY-anisotropy makes the saturated magnetization (the eigenstate
when all spins in cluster are aligned along the magnetic field) inaccessible
for finite magnetic field magnitude (asymptotical saturation). We
demonstrate all these features on three examples: spin-1/2 dimer,
mixed spin-(1/2,1) dimer, spin-1/2 ring trimer. We consider also the
simplest Ising-Heisenberg chain, the Ising-XYZ diamond chain with
four different g-factors. In the chain model the magnetization curve
has a more complicated and non trivial structure which that for clusters.
\end{abstract}



\maketitle

\section{Introduction}

Magnetization curves of low-dimensional quantum antiferromagnets are
topical issue of current research interest, because they often involve
intriguing features such as magnetization plateaus, jumps, ramps and/or
kinks. The spin-1/2 quantum Heisenberg chain, the spin-1/2 quantum
Ising chain in a transverse field, and the spin-1/2 quantum XX chain
in a transverse field are a few paradigmatic examples of exactly solved
quantum spin chains for which zero-temperature magnetization varies
smoothly with rising magnetic field until the saturation magnetization
is reached \cite{mat,kat,pfe}. Contrary to this, the integer-value
quantum Heisenberg chains (and also many other low-dimensional quantum
antiferromagnets) contain in a zero-temperature magnetization process
remarkable magnetization plateau(s) at rational value(s) of the saturation
magnetization \cite{qp1,qp2}. The intermediate plateaus of Heisenberg
spin chains reflect quantum states of matter with exotic topological
order such as the Haldane phase \cite{hal1,hal2}, whereas their presence
is restricted by quantization condition known as Oshikawa-Yamanaka-Affleck
rule \cite{oya1,oya2}.

On the other hand, it could be generally expected that the antiferromagnetic
Heisenberg spin clusters should always exhibit leastwise one intermediate
plateau before the magnetization jumps to its saturation value \cite{js04,js05,js09,js10}.
This naive expectation follows from the energy spectrum of the quantum
Heisenberg spin clusters, which is composed of a few discrete energy
levels that cannot naturally form a continuous energy band needed
for a smooth variation of the magnetization at zero temperature. At
first sight, this argumentation is consistent with the existence of
at least one plateau and magnetization jump, which bears a close relation
to level crossing caused by the external magnetic field. From this
perspective, the quite natural question arises whether or not intermediate
magnetization plateau(s) can be partially or completely lifted from
zero-temperature magnetization curves of the Heisenberg spin clusters.

Another spin systems which should be noted in the context of the small
quantum spin clusters are the Ising-Heisenberg chains. They are the
one-dimensional spin systems where the small quantum spin clusters
are assembled to the chain by alternating with the Ising spins in
such a way, that the Hamiltonian for the whole system is a sum of
mutually commuting block Hamiltonians. These systems have much in
common with the ``classical'' chains of the Ising spins, as they can
be solved by the same technique and the eigenstates are just the direct
product of the eigenstates of the single block, though, for the more
complicated structure the doubling of unit cell is possible. The magnetization
curves, thus, for the Ising-Heisenberg spin systems share almost all
features with the magnetization curves of the small spin clusters,
but can contain much more intermediate magnetization plateaus. Various
variants of the Ising-Heisenberg chains have been examined: diamond-chain\cite{can06,can09,roj11a,roj12,bel13,lis13,lis14,ana14,tor14,faz14,qid,lis15,abg15,eft15,gao15},
sawtooth chain\cite{oha09,bel10}, orthogonal-dimer chain\cite{oha12,pau13,ver13},
tetrahedral chain\cite{val08,ant09,oha10,roj13a,str14a}, and some
special examples relevant to real magnetic materials\cite{str05,Dy1,Dy2,bel14}.

In the present work, we will rigorously examine a magnetization process
of a few quantum Heisenberg spin clusters and Ising-Heisenberg diamond
chain, which will not display strict magnetization plateaus on assumption
that some constituent spins have different Landé g-factors and may
be a XY-anisotropy of the exchange interaction. Also, the Dzyaloshinskii-Moriya
(DM) term in a direction different from that of the magnetic field
 can lead to the same effect. All those features of the spin
Hamiltonian make the magnetization non-conserved, i.e. non-commutating
with the Hamiltonian. This specific requirement naturally leads to
a nonlinear dependence of the energy levels on a magnetic field, which
consequently causes a smooth change of the magnetization with the
magnetic field within one and the same eigenstate. Although the
smooth change of magnetization due to a difference in Landé g-factors
or/and XY-anisotropy and non-collinear DM-term may be quite reminiscent
of that of quantum spin chains with continuous energy bands, it is
of course of completely different mechanism with much simpler origin.

The single-chain magnet, $[\{(\text{CuL})_{2}\text{Dy}\}\{\text{Mo}(\text{CN})_{8}\}]\cdot\text{2CH}_{3}\text{CN}\cdot\text{H}_{2}\text{O}$\cite{Dy1,Dy2,bel14}
is a remarkable example of both Ising-Heisenberg one-dimensional spin
system and a spin model with different Landé g-factors, leading to
non-plateau form of the region of the magnetization curve corresponding
to the same eigenstate. However, as the exact analysis shows\cite{bel14}
the effect is just hardly visible in the magnetization curve plot
in virtue of the very small difference in Landé g-factors of the magnetic
ions, though, the exact expression for the magnetization has explicit dependence
on the magnetic field. Almost the same effect but even quantitatively less pronounced have been observed in the approximate model of the one-dimensional
magnet, the F-F-AF-AF spin chain compound Cu(3-Chloropyridine)$_{2}$(N$_{3}$)$_{2}$\cite{str05}.

The organization of this paper is as follows. In the next Section,
we will clarify a few general statements closely related to an absence
of actual plateaus in zero-temperature magnetization curves of quantum
spin clusters and chains. These arguments of general validity will
be subsequently illustrated on a few specific examples of the spin-1/2
quantum Heisenberg dimer, the mixed spin-(1/2,1) Heisenberg dimer,
the spin-1/2 Heisenberg trimer and the spin-1/2 Ising-Heisenberg diamond
chain in the following four Sections. The summary of the most important
findings along with the implications for experimental systems will
be presented in the concluding part.

\section{General statements}

Let us first start with a few very general statements elucidating
the issue of the non-constant magnetization within one physical state
or the explicit magnetic field dependence of the magnetization corresponding
to a certain eigenstate of the small spin clusters. Obviously, the
aforementioned phenomenon arises when the magnetization is not a good
quantum number
\begin{eqnarray}
\left[\mathcal{H},\;\mathcal{M}^{z}\right]\neq0,\label{comHM}
\end{eqnarray}
Here, $\mathcal{H}$ stands for the Hamiltonian of a spin cluster.
One can distinguish two cases, when $z$-projection of the total spin
$S_{tot}^{z}$ does not commute with the Hamiltonian and the magnetization
operator is proportional to it
\begin{eqnarray}
\left[\mathcal{H},\;S_{tot}^{z}\right]\neq0,\;\;\;\mathcal{M}^{z}=g\mu_{B}S_{tot}^{z},\label{prop}
\end{eqnarray}
or when the $z$-projection of the total spin $S_{tot}^{z}$ is a
good quantum number, but the magnetization operator is not proportional
to it and does not commute with the Hamiltonian
\begin{eqnarray}
\left[\mathcal{H},\;S_{tot}^{z}\right]=0,\;\;\;\mathcal{M}^{z}\neq g\mu_{B}S_{tot}^{z}.\label{nonprop}
\end{eqnarray}
Of course, another possibility is to have the magnetization which
is non proportional to $S_{tot}^{z}$ and the $z$-projection of the
total spin $S_{tot}^{z}$ non-conserved. The spin Hamiltonians, which
do not commute with $S_{tot}^{z}$, usually contain XY-anisotropy
or/and DM vector with a non-zero X or Y part. The magnetization is
non-proportional to the total spin $S_{tot}^{z}$ when the spins possess
different Landé g-factors.

\section{Spin-1/2 Heisenberg dimer}

In this section we consider the spin-1/2 Heisenberg dimer as the simplest
system of two interacting quantum spins described by the most general
Hamiltonian
\begin{eqnarray}
\mathcal{H}_{dim} & = & J\left\{ \left(1+\gamma\right)S_{1}^{x}S_{2}^{x}+\left(1-\gamma\right)S_{1}^{y}S_{2}^{y}+\Delta S_{1}^{z}S_{2}^{z}\right\} \nonumber \\
 &  & +\mathbf{D}\cdot\left(\mathbf{S}_{1}\times\;\mathbf{S}_{2}\right)-\mathbf{B}\cdot\left(g_{1}\bm{S}_{1}+g_{2}\bm{S}_{2}\right).\label{ham2}
\end{eqnarray}
Here, $S_{1,2}^{\alpha}$, $(\alpha=x,y,z)$ are the spatial components
of the spin-1/2 operators for two spins in the dimer. We assume the
fully anisotropic XYZ Heisenberg coupling with two anisotropy constants
$\gamma$, $\Delta$ and two different but isotropic Landé g-factors.
The spatial direction of the magnetic field $\bm{B}$ and the DM-vector
$\bm{D}$ are arbitrary so far. Without loss of generality, one may
however choose a direction of the magnetic field along the z-axis
and the DM vector to lie in xz-plane
\begin{eqnarray}
\mathcal{H}_{dim} & = & J\left\{ \left(1+\gamma\right)S_{1}^{x}S_{2}^{x}+\left(1-\gamma\right)S_{1}^{y}S_{2}^{y}+\Delta S_{1}^{z}S_{2}^{z}\right\} \nonumber \\
 &  & +D_{x}\left(S_{1}^{y}S_{2}^{z}-S_{1}^{z}S_{2}^{y}\right)+D_{z}\left(S_{1}^{x}S_{2}^{y}-S_{1}^{y}S_{2}^{x}\right)\nonumber \\
 &  & -B\left(g_{1}S_{1}^{z}+g_{2}S_{2}^{z}\right).\label{ham2a}
\end{eqnarray}
Let us calculate the commutators of the Hamiltonian (\ref{ham2a})
with the z-projections of the operators corresponding to the total
spin and magnetization
\begin{eqnarray}
 &  & S_{tot}^{z}=S_{1}^{z}+S_{2}^{z},\label{SM}\\
 &  & \mathcal{M}^{z}=g_{1}S_{1}^{z}+g_{2}S_{2}^{z},\nonumber
\end{eqnarray}
\begin{alignat}{1}
\left[\mathcal{H}_{dim},S^{z}\right]= & -2i\gamma\left(S_{1}^{x}S_{2}^{y}+S_{1}^{y}S_{2}^{x}\right)+iD_{x}\left(S_{1}^{x}S_{2}^{z}-S_{1}^{z}S_{2}^{x}\right),\nonumber \\
\left[\mathcal{H}_{dim},\mathcal{M}^{z}\right]= & -ig_{-}D_{z}\left(S_{1}^{x}S_{2}^{x}+S_{1}^{y}S_{2}^{y}\right)+\nonumber \\
 & +ig_{-}J\left(S_{1}^{x}S_{2}^{y}-S_{1}^{y}S_{2}^{x}\right)+\nonumber \\
 & -i\gamma g_{+}\left(S_{1}^{x}S_{2}^{y}+S_{1}^{y}S_{2}^{x}\right)+\nonumber \\
 & +iD_{x}\left(g_{1}S_{1}^{x}S_{2}^{z}-g_{2}S_{1}^{z}S_{2}^{x}\right),
\end{alignat}
where $g_{\pm}=g_{1}\pm g_{2}$. As one can see, the XY-anisotropy,
$\gamma$, and the DM-vector x-projection, $D_{x}$, make the $S_{tot}^{z}$
and $\mathcal{M}^{z}$ non conserved, but even if we set them to zero,
the magnetization may still be a non-conserved quantity because of
the difference in Landé g-factors. Thus, the spin-1/2 Heisenberg dimer
may exhibit the non-constant magnetization within one ground state
if at least one of the parameters, $g_{2}-g_{1}$, $\gamma$ or $D_{x}(D_{y})$
is non-zero. Let us put $D_{x}=0$ as it makes the analytic calculations
quite cumbersome (the eigenvalue problem leads to a quartic equation), and start with the exact diagonalization of the
Hamiltonian for the anisotropic spin-1/2 Heisenberg dimer with different
Landé g-factors. The eigenvalues are
\begin{eqnarray}
 &  & \varepsilon_{1,2}=-\frac{J\Delta}{4}\pm\frac{1}{2}\sqrt{B^{2}g_{-}^{2}+J^{2}+D_{z}^{2}},\nonumber \\
 &  & \varepsilon_{3,4}=\frac{J\Delta}{4}\pm\frac{1}{2}\sqrt{B^{2}g_{+}^{2}+J^{2}\gamma^{2}}.\label{ev1}
\end{eqnarray}
The corresponding eigenvectors are

\begin{alignat}{1}\label{dim_es}
|\Psi_{1,2}\rangle= & \frac{1}{\sqrt{1+|A_{\pm}|^{2}}}\left(|\uparrow\downarrow\rangle+A_{\pm}|\downarrow\uparrow\rangle\right),\\
A_{\pm}= & \rho_{\pm}e^{i\phi},\;\;\phi=\arctan\frac{D_{z}}{J},\nonumber \\
\rho_{\pm}= & \frac{Bg_{-}\pm\sqrt{B^{2}g_{-}^{2}+J^{2}+D_{z}^{2}}}{\sqrt{J^{2}+D_{z}^{2}}},\nonumber \\
|\Psi_{3,4}\rangle= & \frac{1}{\sqrt{1+B_{\pm}^{2}}}\left(|\uparrow\uparrow\rangle+B_{\pm}|\downarrow\downarrow\rangle\right),\nonumber\\
B_{\pm}= & \frac{Bg_{+}\pm\sqrt{B^{2}g_{+}^{2}+J^{2}\gamma^{2}}}{J\gamma}.\nonumber
\end{alignat}
Under the conditions $g_{1}=g_{2}$ (and $D_{z}=0$) the first two
eigenstates become conventional singlet and $S_{tot}^{z}=0$ component
of the triplet respectively. However, there is no continuous transition
to the $S_{tot}^{z}=1$ and $S_{tot}^{z}=-1$ components of the triplet
in $|\Psi_{3,4}\rangle$ at $\gamma\rightarrow0$. In order to obtain
$|\uparrow\uparrow\rangle$ and $|\downarrow\downarrow\rangle$ one
has to put $\gamma=0$ in the Hamiltonian before diagonalization.
Let us calculate the magnetization eigenvalues for all those eigenstates:
\begin{alignat}{1}
\langle\Psi_{1,2}|\left(g_{1}S_{1}^{z}+g_{2}S_{2}^{z}\right)|\Psi_{1,2}\rangle & =\frac{1}{2}g_{-}\frac{1-\rho_{\pm}^{2}}{1+\rho_{\pm}^{2}}\nonumber \\
 & =\mp\tfrac{Bg_{-}^{2}}{2\sqrt{B^{2}g_{-}^{2}+J^{2}+D_{z}^{2}}}.\label{mag12}
\end{alignat}
At $g_{1}=g_{2}$ this expression becomes 0. However, at $g_{1}\neq g_{2}$
we have explicit dependence of the eigenvalue, corresponding to the
certain eigenstate on the magnetic field. This leads to a non constant
magnetization for the given eigenstate. Thus, we have here $S_{tot}^z=0$ and $\mathcal{M}^z\neq 0$.
\begin{figure}
\centering{}\includegraphics[width=1\columnwidth]{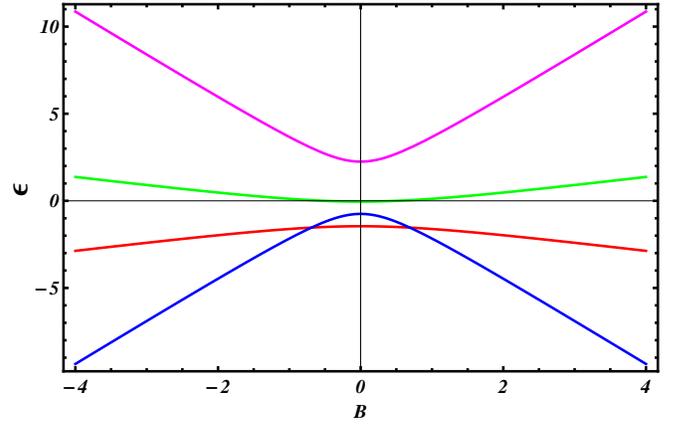} \protect\caption{The energy spectrum of the isolated $S$=1/2 dimer with $g_{1}=2$, $g_{2}=3$, $J=1$, $D_{z}=1$, $\gamma=2$ and $\Delta=3$ displaying level crossing. Two bottom curves corresponds to $\varepsilon_{2}$ and $\varepsilon_{4}$.
The non-linearity in $B$ on the energy levels is the main reason
for the non-plateau magnetization.}
\label{fig1}
\end{figure}

\begin{figure}
\centering{}\includegraphics[width=1\columnwidth]{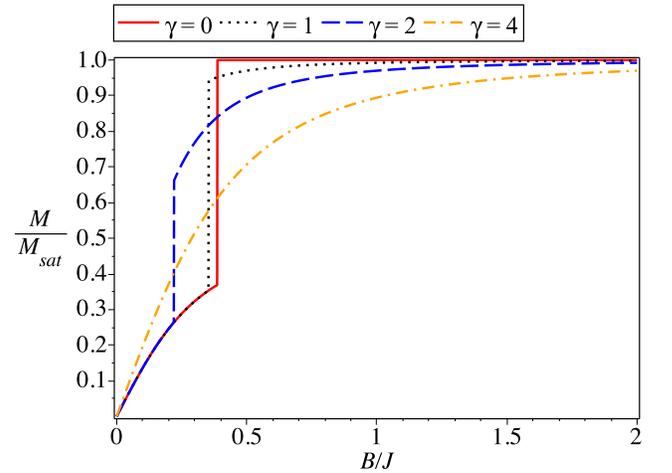} \protect\caption{The zero temperature magnetization curves for the $S$=1/2 dimer
with $g_{1}=2$, $g_{2}=6$, $J=1$, $D_{z}=1$, $\Delta=2$ and $\gamma=0$
(red solid); $\gamma=2$ (black dotted); $\gamma=4$ (blue dashed)
and $\gamma=6$ (brown dot-dashed). $M_{sat}=\frac{1}{2}(g_{1}+g_{2})=4$.}
\label{fig2}
\end{figure}

For the other two eigenstates we have
\begin{alignat}{1}
\langle\Psi_{3,4}|\left(g_{1}S_{1}^{z}+g_{2}S_{2}^{z}\right)|\Psi_{3,4}\rangle & =\frac{1}{2}g_{+}\frac{1-B_{\pm}^{2}}{1+B_{\pm}^{2}}\nonumber \\
 & =\mp\tfrac{Bg_{+}^{2}}{2\sqrt{B^{2}g_{+}^{2}+J^{2}\gamma^{2}}}.\label{mag34}
\end{alignat}
At $\gamma=0$ the expression transform to
\begin{eqnarray}
\langle\Psi_{3,4}|\left(g_{1}S_{1}^{z}+g_{2}S_{2}^{z}\right)|\Psi_{3,4}\rangle=\mp\frac{1}{2}g_{+},
\end{eqnarray}
which corresponds to $|\uparrow\uparrow\rangle$ and $|\downarrow\downarrow\rangle$
eigenstates. The Eq. (\ref{mag34}) has another important feature.
The transverse quantum fluctuations enhanced by the XY-anisotropy
$\gamma$ reduce the magnetization in z-direction in such a way that
it never reaches its saturated values $\pm\frac{1}{2}(g_{1}+g_{2})$
at any non-zero $\gamma$ and finite magnetic field $B$. It is also
important that even at the equal g-factors, $g_{2}=g_{1}=g$ the magnetization
expectation values for the eigenstates $|\Psi_{3,4}\rangle$ exhibit
explicit magnetic field dependence and do not reach their saturated
values at non-zero $\gamma$. Another case of interest is the $g_{2}=-g_{1}=g$,
when the magnetization expectation value for the $|\Psi_{1,2}\rangle$
is non-zero and exhibits explicit dependence on the magnetic filed
and the eigenstates $|\Psi_{3,4}\rangle$ demonstrate zero magnetization.
Moreover, the corresponding expectation values become singular at
$\gamma=0$, because, as was mentioned above, there is no continuous
limit $\gamma\to0$ in terms of eigenvalues and eigenvectors. The
magnetic susceptibility for the aforementioned states can be obtained
in a straightforward way by taking a derivative of the Eqs. (\ref{mag12})
and (\ref{mag34}) with respect to $B$:

\begin{alignat}{1}
\frac{\partial}{\partial B}\langle\Psi_{1,2}|\left(g_{1}S_{1}^{z}+g_{2}S_{2}^{z}\right)|\Psi_{1,2}\rangle= & \tfrac{J^{2}g_{-}^{2}}{2\sqrt{\left(B^{2}g_{-}^{2}+J^{2}\right)^{3}}},\label{chi1234}\\
\frac{\partial}{\partial B}\langle\Psi_{3,4}|\left(g_{1}S_{1}^{z}+g_{2}S_{2}^{z}\right)|\Psi_{3,4}\rangle= & \tfrac{J^{2}\gamma^{2}g_{+}^{2}}{2\sqrt{\left(B^{2}g_{+}^{2}+J^{2}\gamma^{2}\right)^{3}}}.\nonumber
\end{alignat}

The deviation form the horizontal line for the zero-temperature magnetization
curve of the dimer under consideration, thus, is governed by three
factors. Difference of Landé g-factors, $g_{2}-g_{1}$ is responsible
for the non-plateau behavior for the initial part of the magnetization
curve, $B\leq B_{c}$, the larger is the absolute value of the difference
the more pronounced the deviation is. At the same time, the overall
Landé g-factor, $g_{1}+g_{2}$ and the XY-anisotropy $\gamma$ make
another part of the magnetization curve, which in the limit $\gamma=0$
correspond to the saturation, $|\uparrow\uparrow\rangle$, non flat.
The critical field $B_{c}$ is found from the level crossing. As for
the zero temperature only $|\Psi_{2}\rangle$ and $|\Psi_{4}\rangle$
are realized we can found the corresponding value of the magnetic
field from the equation, $\varepsilon_{2}=\varepsilon_{4}$, which
leads to

\begin{widetext}
\begin{eqnarray}
 &  & B_{c}=\frac{\sqrt{2g_{1}g_{2}\Gamma_{-}+J\Delta\left\{ (g_{1}^{2}+g_{2}^{2})J\Delta+\sqrt{4g_{1}^{2}g_{2}^{2}\left((g_{1}^{2}+g_{2}^{2})\Gamma_{-}+2g_{1}g_{2}\Gamma_{+}\right)+(g_{1}^{2}-g_{2}^{2})^{2}J^{2}\Delta^{2}}\right\} }}{2\sqrt{2}g_{1}g_{2}},\label{Bc}
\end{eqnarray}

\end{widetext} with
\begin{equation}
\Gamma_{\pm}=(D_{z}^{2}+J^{2}(1\pm\gamma^{2})).
\end{equation}

\begin{figure}[b]
\centering{} \includegraphics[width=1\columnwidth]{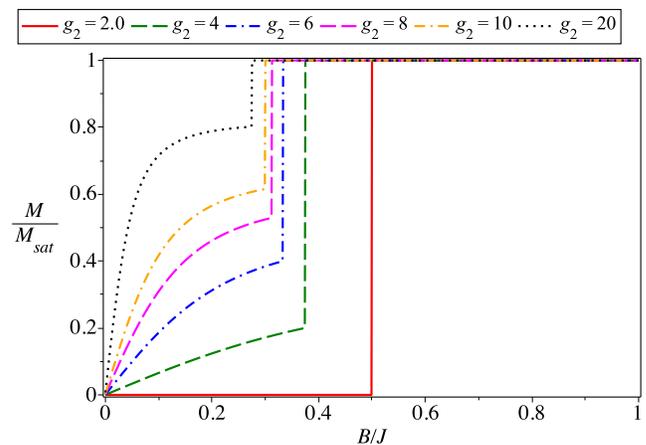}\protect\caption{ The normalized zero-temperature magnetization curves, $M/M_{sat}$
of the $S=1/2$ dimer with two different $g$-factors in case
of isotropic exchange interaction, $\gamma=0$ and $\Delta=1$ . Here,
for the sake of simplicity, we put $J=1,D_{z}=0,g_{1}=2$ and present
the curves for the different values of $g_{2}$. From the bottom to
top $g_{2}$=2 (red); 4 (green); 6 (blue); 8 (magenta); 10 (brown)
and 20 (black). $M_{sat}=\frac{1}{2}(g_{1}+g_{2})$.}
\label{fig3}
\end{figure}

The typical picture of the level crossing curve one can see in Fig.
\ref{fig1}. However, the $B=0$ ground state is also affected by
the value of the XY-anisotropy $\gamma$. The ground state becomes $|\Psi_{4}\rangle$
for a sufficiently large $\gamma$ above certain critical value $\gamma_c$. The
critical value is given by the equation
\begin{eqnarray}
\gamma_{c}=\Delta+\sqrt{1+(D_{z}/J)^{2}}.\label{gamcr}
\end{eqnarray}
The non-linear behavior with respect to the magnetic field is the
main reason for the non-plateau magnetization. As the DM-term in z-direction
does not bring any qualitatively new physics, we can hereafter put
$D_{z}=0$. The expression for the critical field (\ref{Bc}) does
not lead to a proper $\gamma=0$ limit. The case of isotropic Heisenberg
interaction must be considered separately. In the case of isotropic
Heisenberg interaction the difference is only in the saturated states
presented here, and which transforms to $|\Psi_{3,4}\rangle$ at non-zero
$\gamma$, while the $|\Psi_{1,2}\rangle$ eigenstates remain the same.
The value of critical field in this case is
\begin{eqnarray}
B_{c}=J\frac{g_{+}\Delta+\sqrt{g_{-}^{2}\Delta^{2}+4g_{1}g_{2}}}{4g_{1}g_{2}}.\label{Bc0}
\end{eqnarray}
Thus, the jump to the saturated magnetization takes place for the $\gamma=0$
at this value of the magnetic field. The magnitude of the jump depends on the difference of the Landé g-factors and is
given by
\begin{equation}
\Delta M=\frac{g_{+}}{2}\left\{ 1-\tfrac{g_{-}^{2}\left[\Delta+\frac{\sqrt{g_{-}^{2}\Delta^{2}+4g_{1}g_{2}}}{g_{+}}\right]}{4g_{1}g_{2}\sqrt{1+\frac{g_{-}^{2}\left[g_{+}\Delta+\sqrt{g_{-}^{2}\Delta^{2}+4g_{1}g_{2}}\right]}{16g_{1}^{2}g_{2}^{2}}}}\right\} .\label{DeltM}
\end{equation}
The corresponding plots of the zero-temperature magnetization one
can find in the Figs. \ref{fig2} and \ref{fig3}. In the Fig.
\ref{fig2} the evolution of the $T=0$ ground state for different
values of $\gamma$ are presented for $J=1$, $D_{z}=1$, $\Delta=2$,
$g_{1}=2$ and $g_{2}=6$. The critical value of $\gamma$ at which
the $B=0$ ground state of the spin-1/2 spin dimer changes from $|\Psi_{2}\rangle$
to $|\Psi_{4}\rangle$ for these values of $J,D_{z}$ and $\Delta$
is $\gamma_{c}=2+\sqrt{2}\simeq3.41$. Therefore,
for $\gamma=0$ and $\gamma=2$ one can see magnetization curves with
two eigenstates separated by the jump. The non-plateau behavior of
the magnetization for $|\Psi_{2}\rangle$ at $B<B_{c}$ is well visible.
Also, the non-plateau character of the magnetization curve, corresponding
to $|\Psi_{4}\rangle$ is obvious for $\gamma=2$, while for $\gamma=0$
we see ideal plateau at $M=\frac{1}{2}(g_{1}+g_{2})=4$.
This non-plateau behavior and inaccessibility of the saturation are
more pronounced for $\gamma=4$ and $\gamma=6$ when the system for
all values of the magnetic field ($B>0$) is in the $|\Psi_{4}\rangle$
eigenstate and its $T=0$ magnetization curve demonstrates the form
very similar to the one of a system with a band of magnetic excitations
or/and to the high-temperature magnetization curve given by the Brillouin
function. The effect of the difference between the Landé g-factor
is summarized in Fig. \ref{fig3}. To demonstrate the evolution
of the ground state $|\Psi_{2}\rangle$ under the change of the difference
of Landé g-factors, we  have chosen $\gamma=0$ and $J=1,D_{z}=0,\Delta=1$
and plotted the normalized magnetization $M/M_{sat}$, as saturation
magnetization, $M_{sat}=\frac{1}{2}(g_{1}+g_{2})$ is different for
each curve. For the $g_{1}=g_{2}$ there are just two ideal plateaus
at $M=0$ (singlet state) and $M=1$. The magnetization jumps from
$0$ to $1$ at $B_{c}=J\frac{1+\Delta}{2g}$ (here $g_{1}=g_{2}=g$).
However, the essential changes appear when the difference between
$g$-factors is growing. For $g_{2}-g_{1}$ non equal to zero the part
of the magnetization curve corresponding to the $S_{1}^{z}+S_{2}^{z}=0$
deviates from the horizontal line and become almost linear (for small
$g_{-}$) and then more and more rapidly growing with the shift of
the transition point between $|\Psi_{2}\rangle$ and $|\Psi_{4}\rangle$
in the lower $B$ region.

  Of course, the effects of the DM-terms in molecular magnets and other low-dimensional many-body spin systems have been intensively studied in various contexts during last decade\cite{DM1,DM2,DM3,DM4,DM5}. In Ref. \onlinecite{DM2}  the isolated spin-dimer with DM-terms has been considered with general mutual orientation of the DM-vector and magnetic field,
\begin{eqnarray}
\mathcal{H}=J\;\mathbf{S}_1\cdot \mathbf{S}_2+\mathbf{D}\cdot\left(\mathbf{S}_1\times\mathbf{S}_2\right)-g B (S_1^z+S_2^z),\label{H2DM}
\end{eqnarray}
where $\mathbf{D}=\left(0, D\sin\theta,D\cos\theta\right)$. Though, as was shown above, in this case the eigenvalue problem leads to the solution of quartic equation, the authors found an approximate ground state in the limit $D/J\ll 1$ and below the critical field $B=J/g$. They explicitly found out the magnetization of the ground state which turned out to be linear in $B$
\begin{eqnarray}
\mathbf{M}=\frac{g}{4J^3}\left(\mathbf{D}\times\mathbf{B}\right)\times\mathbf{D}.
\end{eqnarray}
This is the approximate form of the non-linear behavior of the magnetization we have obtained exactly above in the case of $\mathbf{D}=(0,0,D_z)$. Despite of all these results, the issue of the non-conserving magnetization and its consequences have not been systematically investigated so far.

\begin{figure}[h]
\centering{} \includegraphics[width=1\columnwidth]{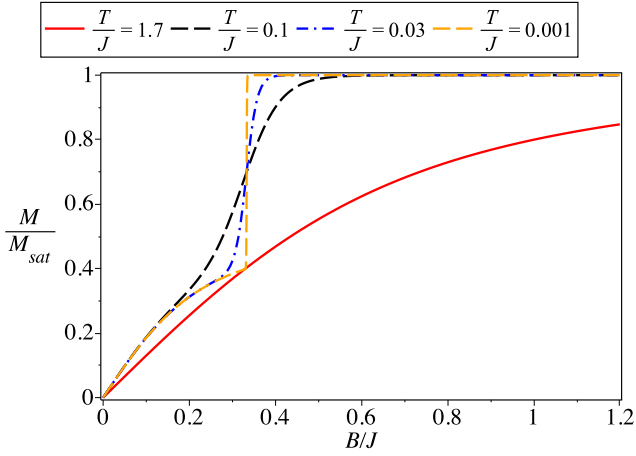} \protect\caption{ The normalized finite-temperature magnetization curves, $M/M_{sat}$
for the $S=1/2$ spin dimer with two different $g$-factors in case
of isotropic exchange interaction, $\gamma=0$ and $\Delta=1$ at
different temperatures for $g_{1}=2$, $g_{2}=6$, $J=1$, $\Delta_{z}=\gamma=0$.
$T/J=1.7$ (red, solid); $T/J=0.1$ (black, dotted); $T/J=0.03$ (blue, dashed)
and $T/J=0.001$ (brown, dot-dashed). $M_{sat}=\frac{1}{2}(g_{1}+g_{2})=4$.}
\label{fig4}
\end{figure}

It is also straightforward to construct the thermodynamics of the
simple dimer. The partition function is calculated directly from the
spectrum:
\begin{alignat}{1}
Z_{dim}= & 2\left\{ e^{\beta\frac{J\Delta}{4}}\cosh\left[\frac{\beta}{2}\sqrt{B^{2}g_{-}^{2}+J^{2}+D_{Z}^{2}}\right]+\right.\nonumber \\
 & +\left.e^{-\beta\frac{J\Delta}{4}}\cosh\left[\frac{\beta}{2}\sqrt{B^{2}g_{+}^{2}+J^{2}\gamma^{2}}\right]\right\} .\label{Zdimer}
\end{alignat}
The magnetization is found in a standard way, as $M_{dim}=\frac{1}{\beta}\left(\frac{\partial\log Z_{dim}}{\partial B}\right)_{\beta}$,
yielding
\begin{alignat}{1}
M_{dim}= & \frac{B}{Z_{dim}}\left\{ \tfrac{g_{+}^{2}e^{-\beta\frac{J\Delta}{4}}}{\sqrt{B^{2}g_{-}^{2}+J^{2}\gamma^{2}}}\sinh\left[\tfrac{\beta\sqrt{B^{2}g_{+}^{2}+J^{2}\gamma^{2}}}{2}\right]+\right.\nonumber \\
 & \left.+\tfrac{g_{-}^{2}e^{\beta\frac{J\Delta}{4}}}{\sqrt{B^{2}g_{-}^{2}+J^{2}+D_{Z}^{2}}}\sinh\left[\tfrac{\beta\sqrt{B^{2}g_{-}^{2}+J^{2}+D_{Z}^{2}}}{2}\right]\right\} .
\end{alignat}
The plots of the finite-temperature magnetization for the S=1/2 spin
dimer are presented in Fig. \ref{fig4}. It is worth mentioning that
thermal fluctuations eliminate from magnetization curves all typical structures (such as plateaus
or quasi-plateaus) quite similarly as the large XY-anisotropy does for zero-temperature magnetization curves.

\section{Mixed-spin Heisenberg dimer}

\begin{figure}
\centering{}\includegraphics[width=1\columnwidth]{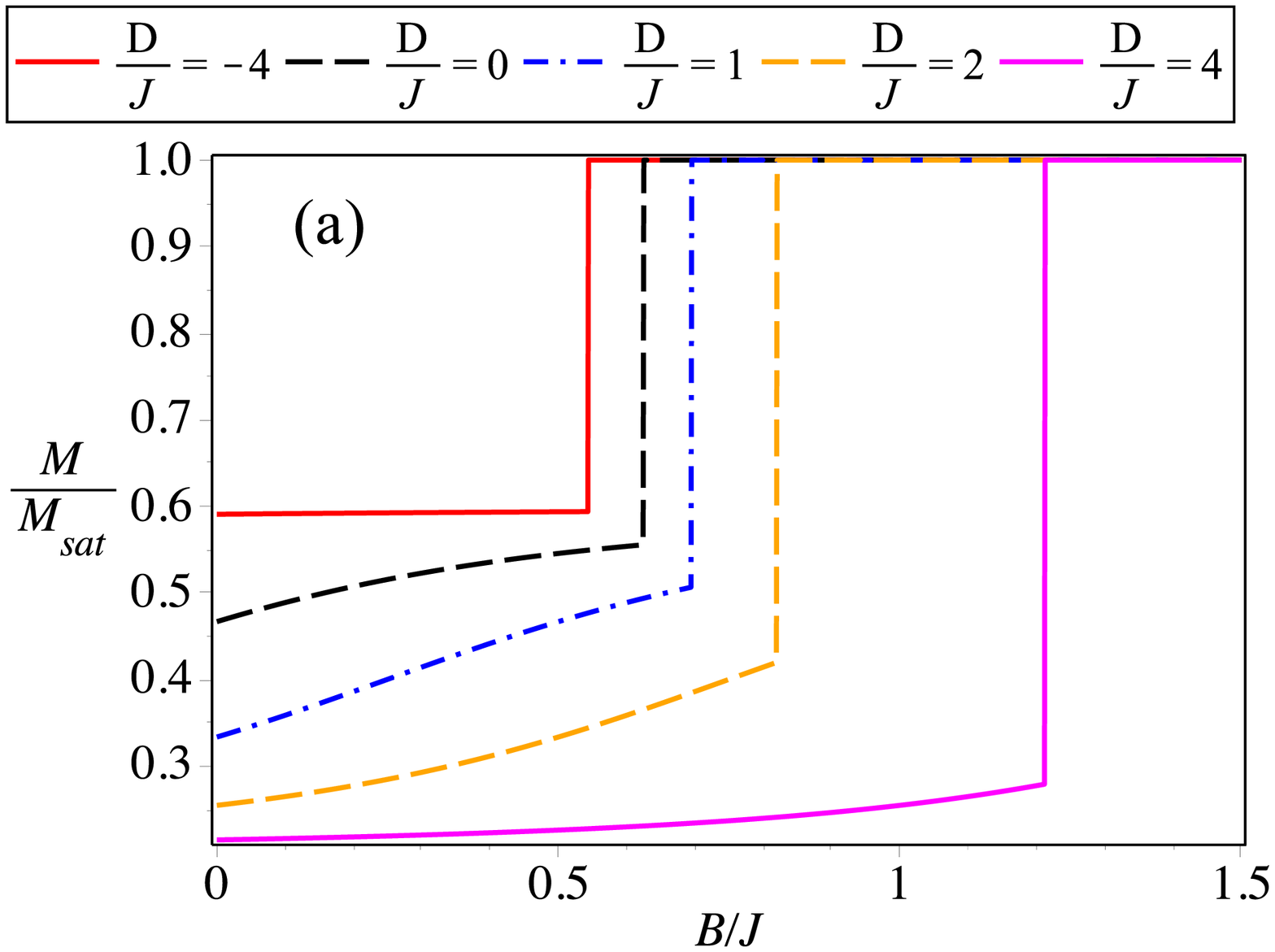} \includegraphics[width=1\columnwidth]{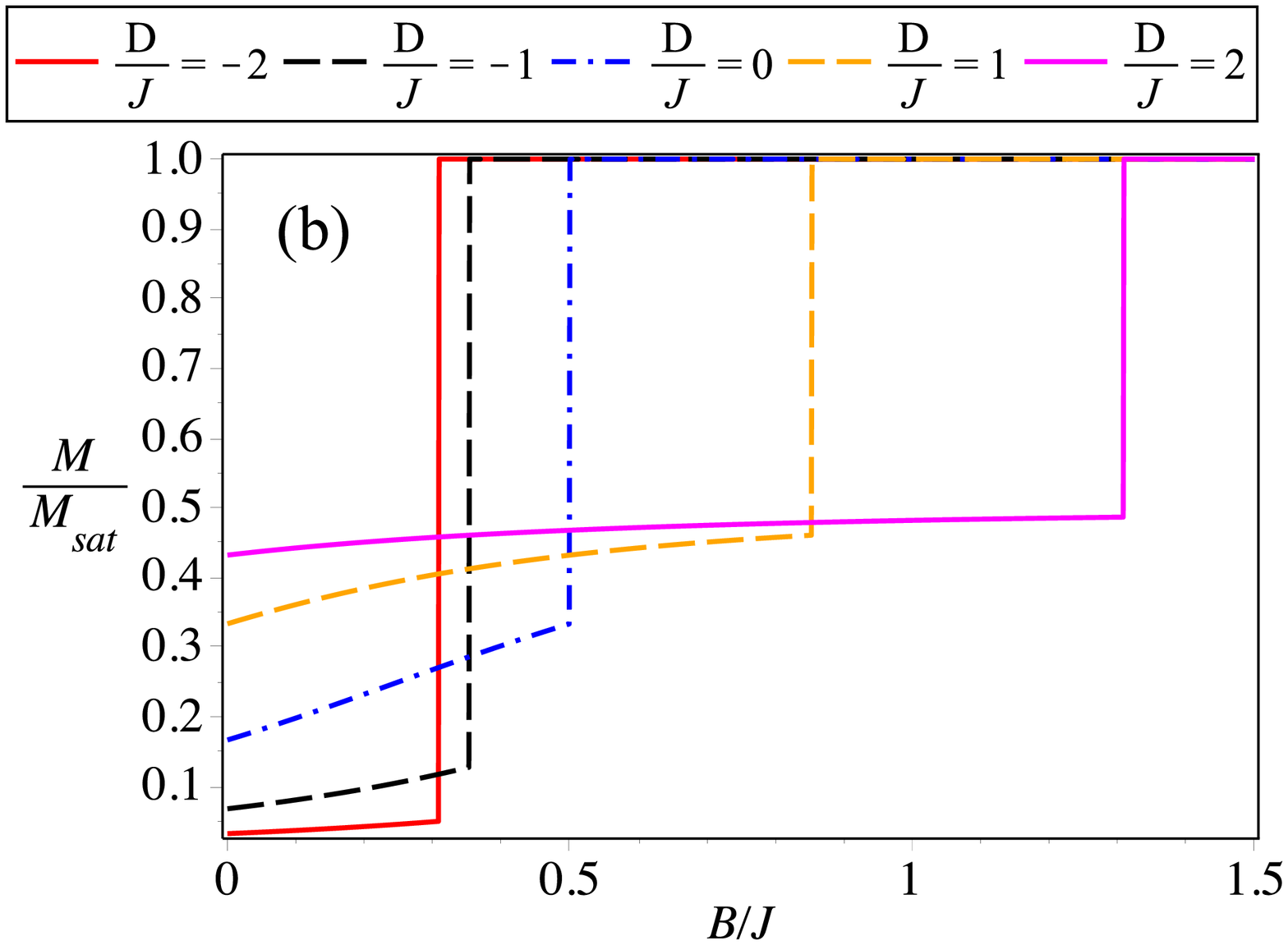}
\protect\caption{\label{figmixab} Zero-temperature normalized magnetization curves
of the mixed spin-(1/2,1) dimer by assuming the isotropic Heisenberg
coupling $\Delta=1$, several values of the uniaxial single-ion anisotropy
and two different sets of the Landé g-factors: (a) $g_{1}=2$, $g_{2}=4$;
(b) $g_{1}=4$, $g_{2}=2$. $M_{sat}=\frac{1}{2}(g_{1}+2g_{2})$.}
\end{figure}

Another interesting example of a simple quantum spin system, which
may possibly show a striking dependence of the total magnetization
on a magnetic field, is the mixed spin-(1/2,1) Heisenberg dimer defined
by the Hamiltonian
\begin{eqnarray}
\mathcal{H}_{mixed} & = & J\left(S_{1}^{x}\mu_{2}^{x}+S_{1}^{y}\mu_{2}^{y}+\Delta S_{1}^{z}\mu_{2}^{z}\right),\nonumber \\
 & + & D(\mu_{2}^{z})^{2}-B\left(g_{1}S_{1}^{z}+g_{2}\mu_{2}^{z}\right).\label{Ham}
\end{eqnarray}
Here, $S_{1}^{\alpha}$ and $\mu_{2}^{\alpha}$ ($\alpha=x,y,z$)
represent spatial components of the spin-1/2 and spin-1 operators,
respectively, the exchange constant $J$ denotes the XXZ Heisenberg
coupling between the spin-1/2 and spin-1 magnetic ions, $\Delta$
is an exchange anisotropy in this interaction, $D$ is an uniaxial
single-ion anisotropy acting on a spin-1 magnetic ion, $g_{1}$ and
$g_{2}$ are Landé $g$-factors of the spin-1/2 and spin-1 magnetic
ions in an external magnetic field $B$. As the effect of the XY-anisotropy
was described in details in the previous Section, here to put it simple,
we assume $\gamma=0$. A straightforward diagonalization of the Hamiltonian
(\ref{Ham}) gives a full spectrum of eigenstates, which can be characterized
by the following set of eigenvalues
\begin{eqnarray}
\varepsilon_{1,2} & = & \frac{1}{2}J\Delta+D\mp\frac{B}{2}\left(g_{1}+2g_{2}\right),\nonumber \\
\varepsilon_{3,4} & = & -\frac{1}{4}(J\Delta-2D+2g_{2}B)\nonumber \\
 &  & \mp\frac{1}{4}\sqrt{(J\Delta-2D-2g_{-}B)^{2}+8J^{2}},\nonumber \\
\varepsilon_{5,6} & = & -\frac{1}{4}(J\Delta-2D-2g_{2}B)\nonumber \\
 &  & \mp\frac{1}{4}\sqrt{(J\Delta-2D+2g_{-}B)^{2}+8J^{2}}\label{ep123456}
\end{eqnarray}
and the corresponding eigenvectors
\begin{eqnarray}
|\Psi_{1,2}\rangle & = & |\mp\tfrac{1}{2},\mp1\rangle,\nonumber \\
|\Psi_{3,4}\rangle & = & c_{1}^{\pm}|-\tfrac{1}{2},1\rangle\mp c_{1}^{\mp}|\tfrac{1}{2},0\rangle,\nonumber \\
|\Psi_{5,6}\rangle & = & c_{2}^{\pm}|\tfrac{1}{2},-1\rangle\mp c_{2}^{\mp}|-\tfrac{1}{2},0\rangle,\label{ev123456}
\end{eqnarray}
where the respective probability amplitudes are given by
\begin{eqnarray}
c_{1}^{\pm} & = & \sqrt{\frac{1}{2}\left[1\pm\tfrac{J\Delta-2D-2g_{-}B}{\sqrt{(J\Delta-2D-2g_{-}B)^{2}+8J^{2}}}\right]},\nonumber \\
c_{2}^{\pm} & = & \sqrt{\frac{1}{2}\left[1\pm\tfrac{J\Delta-2D+2g_{-}B}{\sqrt{(J\Delta-2D+2g_{-}B)^{2}+8J^{2}}}\right]}.\label{pa}
\end{eqnarray}
It should be mentioned that the mixed spin-(1/2,1) Heisenberg dimer
exhibits a strict intermediate plateau at one-third of the saturation
magnetization regardless of uniaxial single-ion anisotropy on assumption
that the Landé g-factors of both constituent magnetic ions are equal
$g_{1}=g_{2}$. If the Landé g-factors are different $g_{1}\neq g_{2}$,
then, the mixed spin-(1/2,1) Heisenberg dimer displays more intriguing
zero-temperature magnetization curves basically affected by a relative
strength of the uniaxial single-ion anisotropy. To illustrate the
case, the total normalized magnetization of the mixed spin-(1/2,1)
dimer is plotted in Fig. \ref{figmixab} against the magnetic field
for the isotropic Heisenberg coupling $\Delta=1$, several values
of the uniaxial single-ion anisotropy and two particular sets of Landé
g-factors. A smooth variation of the total magnetization observed
below a saturation field relates to a gradual change of probability
amplitudes of two entangled microstates $|-1/2,1\rangle$ and $|1/2,0\rangle$
within the eigenstate $|\Psi_{3}\rangle$. In a low-field region with
continuously varying magnetization, mean values of two constituent
spins and the total magnetization can be therefore calculated with
the help of the corresponding lowest-energy eigenvector $|\Psi_{3}\rangle$
given by Eq. (\ref{ev123456})
\begin{eqnarray*}
 &  & \langle\Psi_{3}|S_{1}^{z}|\Psi_{3}\rangle=-\frac{1}{2}\tfrac{J\Delta-2D-2g_{-}B}{\sqrt{(J\Delta-2D-2g_{-}B)^{2}+8J^{2}}},\\
 &  & \langle\Psi_{3}|\mu_{2}^{z}|\Psi_{3}\rangle=\frac{1}{2}\left[1+\tfrac{J\Delta-2D-2g_{-}B}{\sqrt{(J\Delta-2D-2g_{-}B)^{2}+8J^{2}}}\right],\\
 &  & M=\langle\Psi_{3}|\left(g_{1}S_{1}^{z}+g_{2}\mu_{2}^{z}\right)|\Psi_{3}\rangle=\\
 &  & \frac{g_{2}}{2}-\frac{g_{-}}{2}\tfrac{J\Delta-2D-2g_{-}B}{\sqrt{(J\Delta-2D-2g_{-}B)^{2}+8J^{2}}}.
\end{eqnarray*}
At $g_{-}=0$ we have here $M=\frac{g}{2}$ ($g_{1}=g_{2}=g$). It
can be seen from Fig. \ref{figmixab} that the respective field variations
of the total magnetization (normalized by the saturation magnetization
$M_{sat}=\frac{1}{2}(g_{1}+2g_{2})$) depend basically on whether
the Landé g-factor of the spin-1 magnetic ion is greater or smaller
than the g-factor of the spin-1/2 magnetic ion. The total magnetization
is gradually suppressed by an increase in the single-ion anisotropy
in the former case $g_{1}<g_{2}$ (see Fig. \ref{figmixab}a), while
the total magnetization is conversely enhanced by an increase in the
single-ion anisotropy in the latter case $g_{1}>g_{2}$ (see Fig.
\ref{figmixab}b). In general, the total magnetization displays a
considerable dependence on a magnetic field for small enough single-ion
anisotropies $D/J\approx0$, while one recovers a quasi-plateau dependence
with only a subtle variation of the total magnetization in two limiting
cases $D/J\to\pm\infty$ at which the following asymptotic values
are reached
\begin{eqnarray}
\lim_{D/J\to\infty}M/M_{sat} & = & \frac{g_{1}}{2g_{2}+g_{1}},\nonumber \\
\lim_{D/J\to-\infty}M/M_{sat} & = & \frac{2g_{2}-g_{1}}{2g_{2}+g_{1}}.\label{asymlim}
\end{eqnarray}

\begin{figure}
\centering{}\includegraphics[width=1\columnwidth]{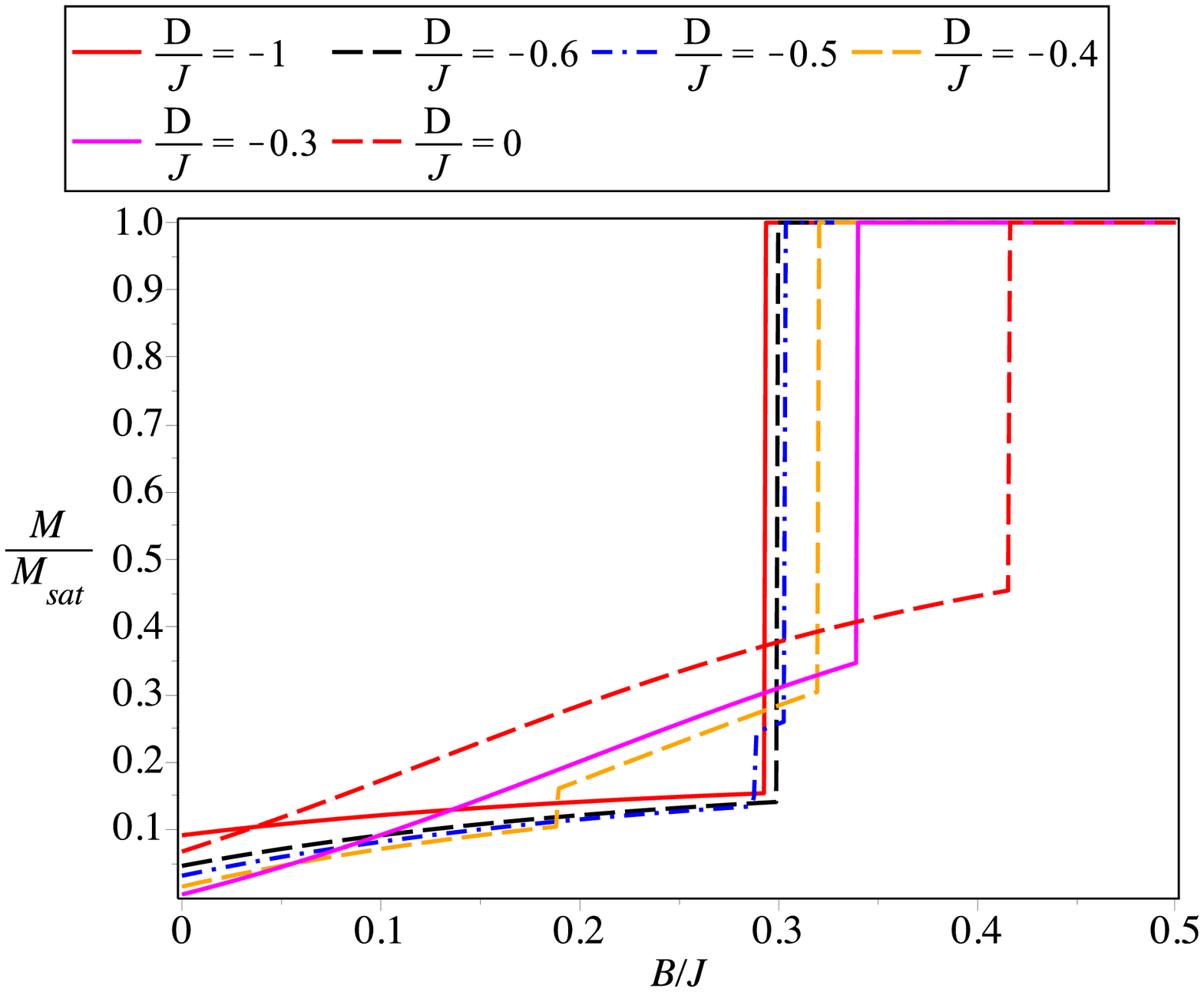} \protect\caption{\label{figmix} Zero-temperature normlized magnetization curves of
the mixed spin-(1/2,1) dimer by assuming the isotropic Heisenberg
coupling $\Delta=1$, several values of the uniaxial single-ion anisotropy
and the Landé g-factors $g_{1}=6$ and $g_{2}=2$. $M_{sat}=\frac{1}{2}(g_{1}+2g_{2})$.}
\label{fig6}
\end{figure}

However, the most surprising zero-temperature dependence of the total
magnetization can be found when the g-factor of the spin-1/2 magnetic
ion is much greater than the g-factor of the spin-1 magnetic ion ($g_{1}\gg g_{2}$)
and the uniaxial single-ion anisotropy is of easy-axis type $D<0$.
It turns out that the eigenvector $|\Psi_{5}\rangle$ characterized
by a quantum entanglement of two microstates $|1/2,-1\rangle$ and
$|-1/2,0\rangle$ may eventually become the lowest-energy eigenstate
with a positive value of the total magnetization in spite of negative
value of the total spin $S_{tot}^{z}=-1/2$. It is quite evident that
a strong enough easy-axis single-ion anisotropy suppresses the occurrence
probability of the microstate $|-1/2,0\rangle$, whereas the other
microstate $|1/2,-1\rangle$ may lead to a positive magnetization
due to much greater the Landé g-factor of the spin-1/2 magnetic ion
$g_{1}\gg g_{2}$ than that of the spin-1 magnetic ion. Mean values
of two constitutent spins and the total magnetization follow from
the corresponding lowest-energy eigenvector $|\Psi_{5}\rangle$ given
by Eq. (\ref{ev123456})
\begin{eqnarray}
 &  & \langle\Psi_{5}|S_{1}^{z}|\Psi_{5}\rangle=\frac{1}{2}\tfrac{J\Delta-2D+2g_{-}B}{\sqrt{(J\Delta-2D+2g_{-}B)^{2}+8J^{2}}},\nonumber \\
 &  & \langle\Psi_{5}|\mu_{2}^{z}|\Psi_{5}\rangle=-\frac{1}{2}\left[1+\tfrac{J\Delta-2D+2g_{-}B}{\sqrt{(J\Delta-2D+2g_{-}B)^{2}+8J^{2}}}\right],\label{Ref:psi5}\\
 &  & M=\langle\Psi_{3}|\left(g_{1}S_{1}^{z}+g_{2}\mu_{2}^{z}\right)|\Psi_{3}\rangle=\nonumber \\
 &  & -\frac{g_{2}}{2}+\frac{g_{-}}{2}\tfrac{J\Delta-2D-2g_{-}B}{\sqrt{(J\Delta-2D-2g_{-}B)^{2}+8J^{2}}}.\label{Ref:psi3}
\end{eqnarray}
In the analogy with the $|\Psi_{3}\rangle$ case, here we get $M=-\frac{g}{2}$
at $g_{1}=g_{2}=g$. The zero-temperature magnetization curves displayed
in Fig. \ref{figmix} afford a convincing proof that the total magnetization
may vary continuously in a low-field region, then it may show an abrupt
jump to an intermediate-field region with another continuously varying
magnetization terminating just at the saturation field (see the magnetization
curves for $D/J=-0.4$ and $-0.5$). In accordance with the previous
argumentation, the total magnetization follows the formula (\ref{Ref:psi5})
in the low-field region attributable to the eigenstate $|\Psi_{5}\rangle$,
while it varies according to Eq. (\ref{Ref:psi3}) in the intermediate-field
region attributable to the eigenstate $|\Psi_{3}\rangle$. The magnetization
part corresponding to the eigenstate $|\Psi_{3}\rangle$ gradually
diminishes as the easy-axis single-ion anisotropy strengthens (i.e.
it becomes more negative) and hence, the total magnetization shows
below a saturation field only a single region with continuously varying
magnetization due to the striking lowest-energy eigenstate $|\Psi_{3}\rangle$
with a negative total spin but a positive total magnetization (see
the magnetization curves for $D/J=-0.6$ and $-1.0$). It is straightforward
to calculate the susceptibility for the eigenstates $|\Psi_{3}\rangle$,
and $|\Psi_{5}\rangle$:
\begin{eqnarray}
 &  & \frac{\partial}{\partial B}\langle\Psi_{3,5}|\left(g_{1}S_{1}^{z}+g_{2}\mu_{2}^{z}\right)|\Psi_{3,5}\rangle\label{sus35}\\
 &  & =\pm2g_{-}^{2}\frac{\left(J\Delta-2D-2g_{-}B\right)^{2}+4J^{2}}{\left[\left(J\Delta-2D-2g_{-}B\right)^{2}+8J^{2}\right]^{3/2}},\nonumber
\end{eqnarray}
which become zero only when $g_{1}=g_{2}$ or $D/J\to\infty$ or $\Delta\to\infty$.
It has been demonstrated that a smooth variation of the total magnetization
at zero temperature within one and the same eigenstate requires a
difference between the Landé g-factors. From this perspective, our
theoretical predictions could be more easily experimentally tested
for the mixed spin-(1/2,1) Heisenberg dimer, which represents plausible
model for heterobimetallic dinuclear complexes naturally having two
unequal Landé g-factors due to two different constituting magnetic
ions. While the quasi-plateau phenomenon should still remain a rather subtle effect
in heterodinuclear complexes composed of Cu$^{2+}$ (spin-1/2) and Ni$^{2+}$ (spin-1) magnetic ions
due to a relatively small difference between the g-factors not exceeding a few percent \cite{nicu,cuni},
it should become much more pronounced in heterodinuclear complexes composed of Co$^{3+}$ (spin-1/2)
and Ni$^{2+}$ (spin-1) magnetic ions having much greater difference between g-factors
(typically $g_{{\rm Co}}\approx 5.9$ and $g_{{\rm Ni}}\approx2.3$)\cite{jomie,carlin,coni}.

\section{Spin-1/2 Heisenberg trimer}

The next by complicity spin system with the different $g$-factors
is the triangle with uniform coupling and with only two $g$-factors
given by the Hamiltonian
\begin{eqnarray}
\mathcal{H}_{trim} & = & J\left(S_{1}^{x}S_{2}^{x}+S_{1}^{y}S_{2}^{y}+\Delta S_{1}^{z}S_{2}^{z}\right.\label{ham3}\\
 & + & S_{1}^{x}S_{3}^{x}+S_{1}^{y}S_{3}^{y}+\Delta S_{1}^{z}S_{3}^{z}\nonumber \\
 & + & \left.S_{2}^{x}S_{3}^{x}+S_{2}^{y}S_{3}^{y}+\Delta S_{2}^{z}S_{3}^{z}\right)\nonumber \\
 & - & g_{1}BS_{1}^{z}-g_{2}B\left(S_{2}^{z}+S_{3}^{z}\right).\nonumber
\end{eqnarray}
The Hamiltonian can be diagonalized in a straightforward way. The
eigenvalues are:
\begin{eqnarray}
 &  & \varepsilon_{1,2}=\frac{3J}{4}\mp\frac{1}{2}\left(g_{1}+2g_{2}\right)B,\label{evH1}\\
 &  & \varepsilon_{3,4}=-\frac{J}{4}\left(2+\Delta\right)\mp\frac{1}{2}g_{1}B,,\nonumber \\
 &  & \varepsilon_{5,6}=\frac{J}{4}\left(1-\Delta\right)\pm Q_{+}+\frac{1}{2}g_{2}B,\nonumber \\
 &  & \varepsilon_{7,8}=\frac{J}{4}\left(1-\Delta\right)\pm Q_{-}-\frac{1}{2}g_{2}B,\nonumber
\end{eqnarray}
where
\begin{eqnarray}
Q_{\pm}=\frac{1}{2}\sqrt{2J^{2}+\left(J\pm g_{-}B\right)^{2}}.\label{Q}
\end{eqnarray}
The eigenvectors are
\begin{eqnarray}
 &  & |\Psi_{1}\rangle=|\uparrow\uparrow\uparrow\rangle,\;\;|\Psi_{2}\rangle=|\downarrow\downarrow\downarrow\rangle,\label{evH3}\\
 &  & |\Psi_{3}\rangle=|\uparrow\rangle_{1}|S\rangle_{23},\;\;|\Psi_{4}\rangle=|\downarrow\rangle_{1}|S\rangle_{23},\nonumber \\
 &  & |\Psi_{5,6}\rangle=\frac{1}{\sqrt{2+c_{\pm}^{2}}}\left(\sqrt{2}|\uparrow\rangle_{1}|T_{0}\rangle_{23}+c_{\pm}|\downarrow\rangle_{1}|T_{+}\rangle_{23}\right),\nonumber \\
 &  & |\Psi_{7,8}\rangle=\frac{1}{\sqrt{2+\bar{c}_{\pm}^{2}}}\left(\sqrt{2}|\downarrow\rangle_{1}|T_{0}\rangle_{23}+\bar{c}_{\pm}|\uparrow\rangle_{1}|T_{-}\rangle_{23}\right),\nonumber
\end{eqnarray}
where the number in the lower right angle of the symbol $|\rangle$
correspond to the certain spin in the triangle, and $|S\rangle$,
$|T_{\pm}\rangle$ and $|T_{0}\rangle$ are spin-singlet and components
of the spin-triplet:
\begin{eqnarray}
 &  & |S\rangle=\frac{1}{\sqrt{2}}\left(|\uparrow\downarrow\rangle-|\downarrow\uparrow\rangle\right),\\
 &  & |T_{+}\rangle=|\uparrow\uparrow\rangle,\nonumber \\
 &  & |T_{-}\rangle=|\downarrow\downarrow\rangle,\nonumber \\
 &  & |T_{0}\rangle=\frac{1}{\sqrt{2}}\left(|\uparrow\downarrow\rangle+|\downarrow\uparrow\rangle\right).\nonumber
\end{eqnarray}
And the coefficients are
\begin{eqnarray}
 &  & c_{\pm}=\frac{-J\Delta+2g_{-}B\pm Q_{+}}{J\Delta},\\
 &  & \bar{c}_{\pm}=\frac{-J\Delta-2g_{-}B\pm Q_{-}}{J\Delta}.\nonumber
\end{eqnarray}
For our purposes the eigenvectors from $|\Psi_{5}\rangle$ to $|\Psi_{8}\rangle$
are of special interest, as they demonstrate the monotonous explicit
dependence of the magnetization on the magnetic filed under constant
value of the projection of the total spin, which is $\pm\frac{1}{2}$
in our case. The expectation value of the magnetic moment for the
eigenstate with the lowest energy among the others ($|\Psi_{8}\rangle$
for positive $B$) is
\begin{eqnarray}
 &  & \langle\Psi_{8}|\left\{ g_{1}S_{1}^{z}+g_{2}\left(S_{2}^{z}+S_{3}^{z}\right)\right\} |\Psi_{8}\rangle\label{Mh3}\\
 &  & =\frac{1}{2}\frac{2g_{1}-\left(g_{1}-2g_{2}\right)\bar{c}_{-}^{2}}{2+\bar{c}_{-}^{2}}.\nonumber
\end{eqnarray}
It is easy to see that for the case of the uniform $g$-factors, $g_{1}=g_{2}=g$
the magnetic moment expectation value become a constant equal to $g/2$.
Though, for the eigenvalues of the magnetization operator the limit
$g_{-}=0$ gives the correct result, this is not the case for the
eigenvectors of the Hamiltonian. Thus, one cannot obtain the standard
basis for the spin trimer by putting $g_{1}=g_{2}$ in the Eqs, (\ref{evH3}).
The susceptibility for the continuous magnetization (\ref{Mh3}) is
given by
\begin{eqnarray}
 &  & \frac{\partial}{\partial B}\langle\Psi_{8}|\left\{ g_{1}S_{1}^{z}+g_{2}\left(S_{2}^{z}+S_{3}^{z}\right)\right\} |\Psi_{8}\rangle=\label{susc3}\\
 &  & =-\frac{4g_{-}\bar{c}_{-}}{(2+\bar{c}_{-}^{2})^{2}}\frac{d\bar{c}_{-}^{2}}{dB}.\nonumber
\end{eqnarray}
The susceptibility becomes zero at $g_{1}=g_{2}$. The partition function
for the spin-trimer under consideration can be obtained in a straightforward
way: \begin{widetext}
\begin{eqnarray}
Z_{trim}=2e^{\beta\frac{J}{4}}\left\{ e^{-\beta J}\cosh\left(\beta\frac{(g_{1}+2g_{2})B}{2}\right)+e^{\beta\frac{J\Delta}{2}}\cosh\left(\beta\frac{g_{1}B}{2}\right)+e^{-\beta\frac{J\Delta}{4}}\left[e^{-\beta\frac{g_{2}B}{2}}\cosh\left(\beta Q_{+}\right)+e^{\beta\frac{g_{2}B}{2}}\cosh\left(\beta Q_{-}\right)\right]\right\} \label{Z3}
\end{eqnarray}
\end{widetext}

\begin{figure}[t]
\centering{} \includegraphics[width=1\columnwidth]{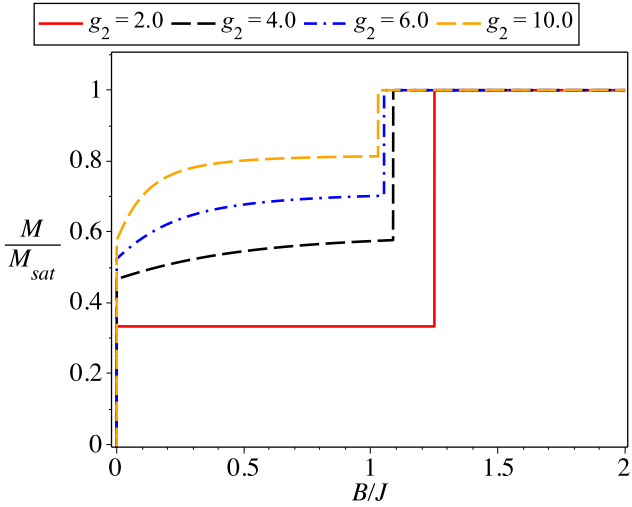} \protect\caption{\label{tr1}The Zero temperature normalized magnetization curves for
the Heisenberg spin trimer with two different Landé g-factors for
$J=1$, $\Delta=2$, $g_{1}=2$ and $g_{2}=2$ (solid red); $g_{2}=4$
(dotted black); $g_{2}=6$ (dashed blue) and $g_{2}=10$ (dot-dashed
brown). $M_{sat}=\frac{1}{2}(g_{1}+2g_{2})$.}
\end{figure}

\begin{figure}[t]
\centering{} \includegraphics[width=1\columnwidth]{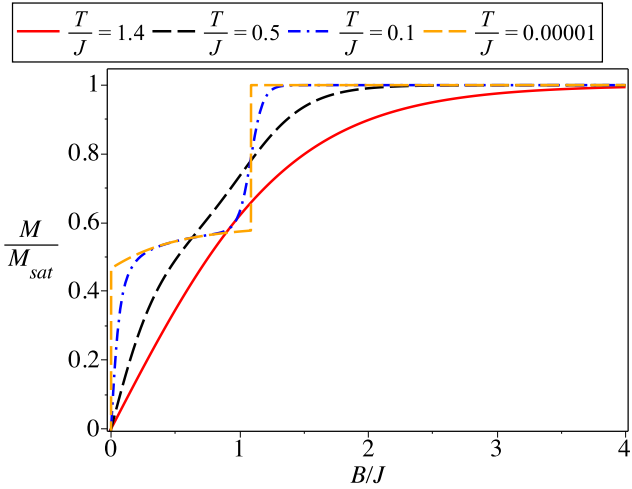} \protect\caption{\label{trT} Finite temperature normalized magnetization curve for
the Heisenberg spin triplet with two different Landé g-factors for
$J=1$, $\Delta=2$, $g_{1}=2$, $g_{2}=4$ and $T=1.4J$ (solid red);
$T=0.5J$ (dotted black); $T=0.1J$ (dashed blue) and $T=0.00001J$
(dot-dashed brown). $M_{sat}=\frac{1}{2}(g_{1}+2g_{2})=5$.}
\end{figure}

The finite-temperature magnetization reads
\begin{widetext}
\begin{eqnarray}
M_{trim} & = & \frac{e^{\beta\frac{J}{4}}}{Z_{tri}}\left\{ (g_{1}+2g_{2})e^{-\beta J}\sinh\left(\beta\frac{(g_{1}+2g_{2})B}{2}\right)+g_{1}e^{\beta\frac{J\Delta}{2}}\sinh\left(\beta\frac{g_{1}B}{2}\right)\right.\label{Mt3}\\
 & + & e^{-\beta\frac{J\Delta}{4}}\left[e^{\beta\frac{g_{2}B}{2}}\left(g_{2}\cosh\left(\beta Q_{-}\right)-\frac{g_{-}(J-2g_{-}B)}{Q_{-}}\sinh\left(\beta Q_{-}\right)\right)\right.\nonumber \\
 & - & \left.\left.e^{-\beta\frac{g_{2}B}{2}}\left(g_{2}\cosh\left(\beta Q_{+}\right)-\frac{g_{-}(J+2g_{-}B)}{Q_{+}}\sinh\left(\beta Q_{+}\right)\right)\right]\right\} \nonumber
\end{eqnarray}
\end{widetext}
The plots of the normalized zero-temperature magnetization
are presented in Fig. \ref{tr1}. Here the development of the non-plateau
part of the magnetization curve with the increase in the difference $g_{2}-g_{1}$
is clearly visible. For comparison the ordinary curve for $g_{1}=g_{2}$
is also presented with a plateau at $M/M_{sat}=1/3$, which corresponds
to the ground state with $S_{tot}^{2}=3/4,S_{tot}^{z}=1/2$:
\begin{eqnarray}
\frac{1}{\sqrt{3}}\left(|\uparrow\uparrow\downarrow\rangle+|\uparrow\downarrow\uparrow\rangle+|\downarrow\uparrow\uparrow\rangle\right),
\end{eqnarray}
which transforms into the $|\Psi_{8}\rangle$ at $g_{2}\neq g_{1}$.
Let us mention also, that the $B=0$ ground state at for $g_{1}=g_{2}$
is four-fold degenerate and the magnetic field lifts this degeneration
just partly, because the 1/3 plateau state is still two-fold degenerate.
The deviation from the horizontal line becomes more pronounced with
the growing difference between g-factors of the spins. Thus, the zero-temperature
magnetization curve for the simple system with finite
discrete spectrum mimics the magnetic behavior of magnets with the
band of magnetic excitations. The value of critical field at which
the level crossing between $|\Psi_{8}\rangle$ and the fully polarized
state $|\Psi_{1}\rangle$  takes place is
\begin{eqnarray}
B_{c}=J\frac{2g_{+}\Delta-g_{1}+\sqrt{(2g_{-}\Delta-g_{1})^{2}+8g_{1}g_{2}}}{4g_{1}g_{2}}.
\end{eqnarray}
In the limit $g_{1}=g_{2}=g$ the value of critical field is $J\frac{1+2\Delta}{2g}$.
The interplay between thermal fluctuations and the non-plateau behavior
can be seen in the Fig. \ref{trT} where one can see  a gradual smearing
out of the magnetization curve with the rise of the
temperature.

\section{Ising-Heisenberg (Ising-XYZ) diamond chain with different Landé g-factors}

\begin{figure}[t]
\centering{} \includegraphics[width=1\columnwidth]{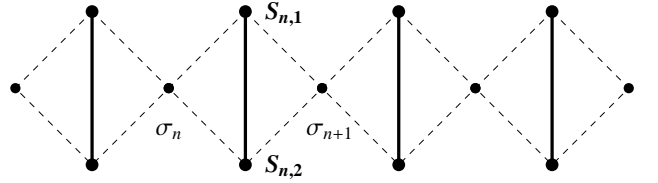} \protect\caption{\label{DC} The Ising-Heisenberg diamond-chain. Solid lines represent
the quantum interactions, while the dotted once stand for the interaction
involving only $z$-components of the spins. Here we also consider
the $g$-factors of the quantum spins $\mathbf{S}_{j,1}$ and $\mathbf{S}_{j,2}$
to be $g_{1}$ and $g_{2}$ respectively.}
\end{figure}

To illustrate the features of having the spin cluster with non-conserving
magnetization as a constituent of the Ising-Heisenberg one-dimensional
systems let us now consider the simplest Ising-Heisenberg spin-chain
with the XYZ-dimers and different g-factors, the diamond chain\cite{can06,can09,roj11a,roj12,bel13,lis13,lis14,ana14,tor14,faz14,qid,lis15,abg15,eft15,gao15}.
However, we are not going to describe the whole problem in all details,
this can be a topic of the forthcoming and separate investigation.
We just want to illustrate how rich the structure of the magnetization
curve can be, if we include the spin cluster with non-conserved magnetization
into the more involved structures. The interest toward the diamond
chain is large not only because of the simplicity of the system, especially
in case of Ising-Heisenberg one-dimensional systems, but also as the
diamond chain is believed to be the real magnetic structure of the
mineral azurite\cite{azu1,azu2,azu3,azu4}. The lattice in depicted
in Fig. \ref{DC}, where the quantum spin-dimer are the vertical
bonds(solid lines) while the dashed lines correspond to Ising couplings.
The Hamiltonian for the whole chain is the sum over the block Hamiltonians:
\begin{alignat}{1}
\mathcal{H}_{dc}= & \sum_{j=1}^{N}\left\{ \mathcal{H}_{j}-\frac{B}{2}\left(g_{j}\sigma_{j}+g_{j+1}\sigma_{j+1}\right)\right\} ,\label{Hamdc1}\\
\mathcal{H}_{j}= & J\left\{ \left(1+\gamma\right)S_{j,1}^{x}S_{j,2}^{x}+\left(1-\gamma\right)S_{j,1}^{y}S_{j,2}^{y}+\Delta S_{j,1}^{z}S_{j,2}^{z}\right\} \nonumber \\
 & +(K(\sigma_{j}+\sigma_{j+1})-g_{1}B)S_{j,1}^{z}\nonumber \\
 & +(K(\sigma_{j}+\sigma_{j+1})-g_{2}B)S_{j,2}^{z},\nonumber
\end{alignat}
where the g-factors of the Ising intermediate spins alternating with
the spin-dimer are also taken alternating
\begin{equation}
g_{j}=\left\{ \begin{array}{cc}
g_{3}, & j\;\mbox{is odd}\\
g_{4}, & j\;\mbox{is even}.
\end{array}\right.
\end{equation}

\subsection{Exact solution}

One can apply the standard technique of the generalized classical
transfer matrix to calculate the free energy of the model under consideration
exactly\cite{roj11a,bel13,tor14,oha09,bel10,oha12,ant09,oha10,bel14}.
However, as here we deal with the alternation of two kind of blocks
one has to compose a two block transfer matrix just by multiplying
transfer matrices for odd and even blocks. In the other words, the
partition function of the model can be factorized in the following
form (the cyclic boundary conditions are assumed):
\begin{alignat}{1}
\mathcal{Z}_{dc} & =\sum_{\sigma_{1},...\sigma_{N}}\;\prod_{j=1}^{N}\;e^{\beta\frac{B}{2}\left(g_{j}\sigma_{j}+g_{j+1}\sigma_{j+1}\right)}\;\mbox{Tr}_{j}\;e^{-\beta\mathcal{H}_{j}\left(\sigma_{j},\sigma_{j+1}\right)}\nonumber \\
 & =\sum_{\sigma_{1},...\sigma_{N}}\;V_{\sigma_{1},\sigma_{2}}\;V_{\sigma_{2},\sigma_{3}}^{T}....V_{\sigma_{N},\sigma_{1}}^{T},
\end{alignat}
where with the aid of the block Hamiltonian eigenvalues from Eq. (\ref{ev1}),
one can obtain four quantities $V_{\sigma_{j},\sigma_{j+1}}$, $\sigma_{j},\sigma_{j+1}=\pm1/2$,
as the entries of the following matrix:
\begin{eqnarray}
\mathbf{V}=2e^{-\beta\frac{J\Delta}{4}}\left(\begin{array}{lc}
e^{\beta\frac{B(g_{3}+g_{4})}{4}}U_{-} & e^{\beta\frac{B(g_{3}-g_{4})}{4}}U_{0}\\
e^{-\beta\frac{B(g_{3}-g_{4})}{4}}U_{0} & e^{-\beta\frac{B(g_{3}+g_{4})}{4}}U_{+}
\end{array}\right),\label{V}
\end{eqnarray}
where
\begin{alignat}{1}
U_{\pm}= & W+\cosh\left(\frac{\beta}{2}\sqrt{\left(Bg_{+}\pm2K\right)^{2}+J^{2}\gamma^{2}}\right),\nonumber \\
U_{0}= & W+\cosh\left(\frac{\beta}{2}\sqrt{B^{2}g_{+}^{2}+J^{2}\gamma^{2}}\right),\nonumber \\
W= & e^{\beta\frac{J\Delta}{2}}\cosh\left(\frac{\beta}{2}\sqrt{B^{2}g_{-}^{2}+J^{2}}\right).
\end{alignat}

Thus, the partition function can be rewritten in the form
\begin{eqnarray}
\mathcal{Z}_{dc}=4^{\frac{N}{2}}e^{-\beta\frac{J\Delta N}{4}}\;\mbox{Tr}\;\mathbf{T}^{\frac{N}{2}},\label{Zdc}
\end{eqnarray}
where,
\begin{widetext}
\begin{eqnarray*}
 &  & \mathbf{T}=\left(\begin{array}{lc}
e^{\beta\frac{B(g_{3}+g_{4})}{2}}U_{-}^{2}+e^{\beta\frac{B(g_{3}-g_{4})}{2}}U_{0}^{2} & \left\{ e^{\beta\frac{Bg_{4}}{2}}U_{-}+e^{-\beta\frac{Bg_{4}}{2}}U_{+}\right\} U_{0}\\
\left\{ e^{\beta\frac{Bg_{4}}{2}}U_{-}+e^{-\beta\frac{Bg_{4}}{2}}U_{+}\right\} U_{0} & e^{-\beta\frac{B(g_{3}+g_{4})}{2}}U_{+}^{2}+e^{-\beta\frac{B(g_{3}-g_{4})}{2}}U_{0}^{2}
\end{array}\right).
\end{eqnarray*}
\end{widetext}

Then, for the free energy per unit cell in the thermodynamic limit,
$N\to\infty$, we have,
\begin{eqnarray}
f=\frac{J\Delta}{4}-\frac{1}{2\beta}\left(\log4+\log\lambda\right),\label{f}
\end{eqnarray}
where the $\lambda$ is the largest eigenvalue of the matrix $\mathbf{T}$,
which is expressed by the entries of the matrix $\mathbf{T}$ in the
following form:
\begin{alignat}{1}
\lambda= & \frac{1}{2}\left(T_{\frac{1}{2},\frac{1}{2}}+T_{-\frac{1}{2},-\frac{1}{2}}+\right.\nonumber \\
 & \left.+\sqrt{\left(T_{\frac{1}{2},\frac{1}{2}}-T_{-\frac{1}{2},-\frac{1}{2}}\right)^{2}+4T_{\frac{1}{2},-\frac{1}{2}}^{2}}\right).\label{lambda}
\end{alignat}

Now, we will analyze the non-plateau magnetization for Ising-XYZ
diamond chain with different g-factors. Using the free energy, one
gets the magnetization
\begin{eqnarray}
M=-\left(\frac{\partial f}{\partial B}\right)_{\beta}=\frac{1}{2\beta}\frac{1}{\lambda}\left(\frac{\partial\lambda}{\partial B}\right)_{\beta}.\label{mag_dc}
\end{eqnarray}
As the lattice has six spins in the translational invariant unit cell,
two $\sigma$ spins and two vertical dimers, the total saturation
magnetization per one block (note that $N$ is the number of block
which is supposed to be even, while the number of the unit cell with
six spins is $N/2$) is
\begin{eqnarray}
M_{sat}=\frac{1}{2}\left(g_{1}+g_{2}+\frac{1}{2}\left(g_{3}+g_{4}\right)\right).\label{mag0_dc}
\end{eqnarray}
One can see the plots of the zero-temperature magnetization of the
Ising-XYZ diamond chain in Fig. (\ref{fig10}). The zero-temperature
curves can be obtained as a sufficiently low-temperature limit of
the exact expression (\ref{mag_dc}). One can see series of quasi-plateau
and magnetization jumps which can undergo a certain variation under
the changing of the XY-anisotropy $\gamma$ (upper panel) or axial
anisotropy $\Delta$ (lower panel).

\subsection{Ground states}

\begin{figure}[h]
\centering{}\label{mag-ch-a} \includegraphics[width=1\columnwidth]{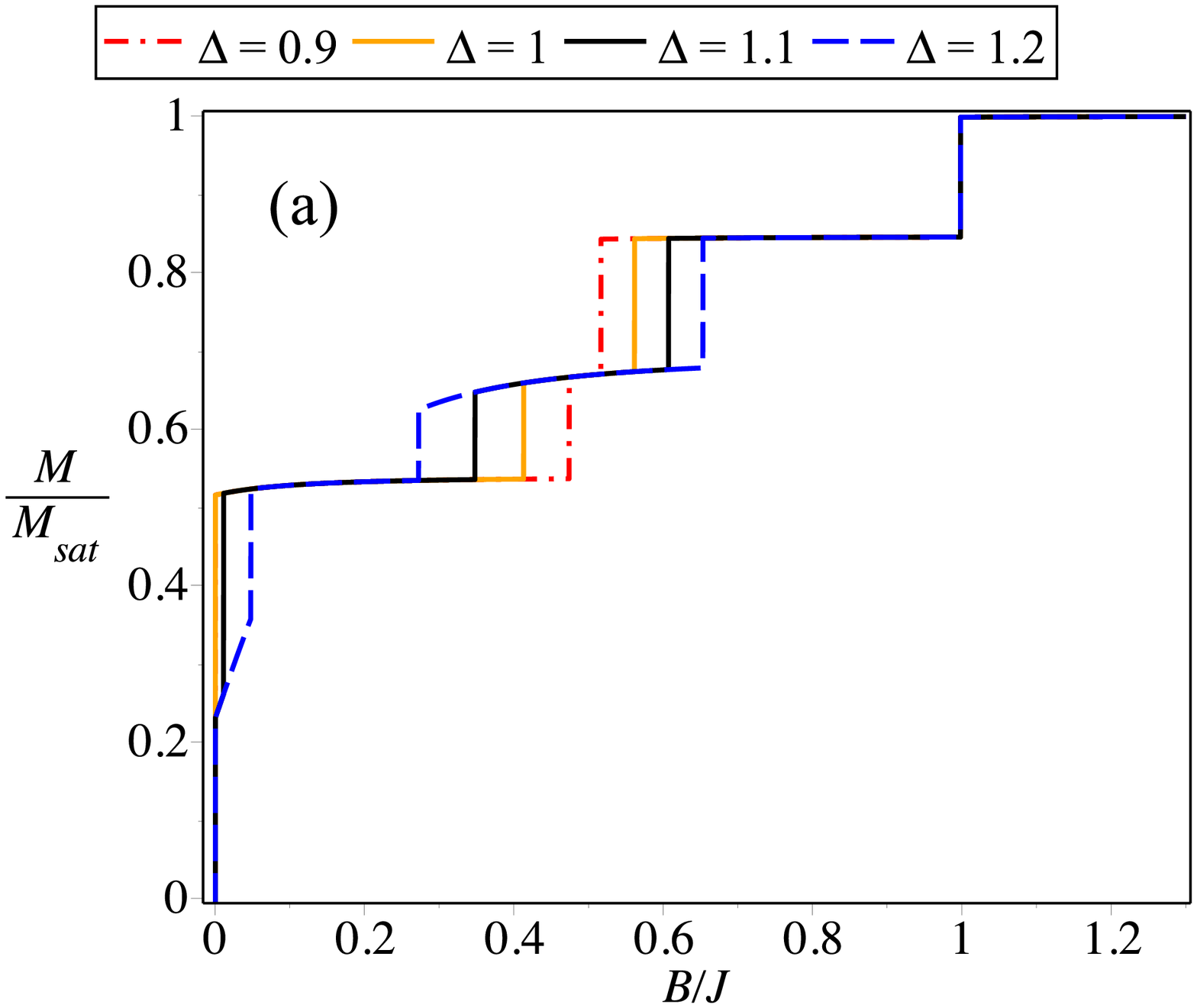}
\centering{}\label{mag-ch-b} \includegraphics[width=1\columnwidth]{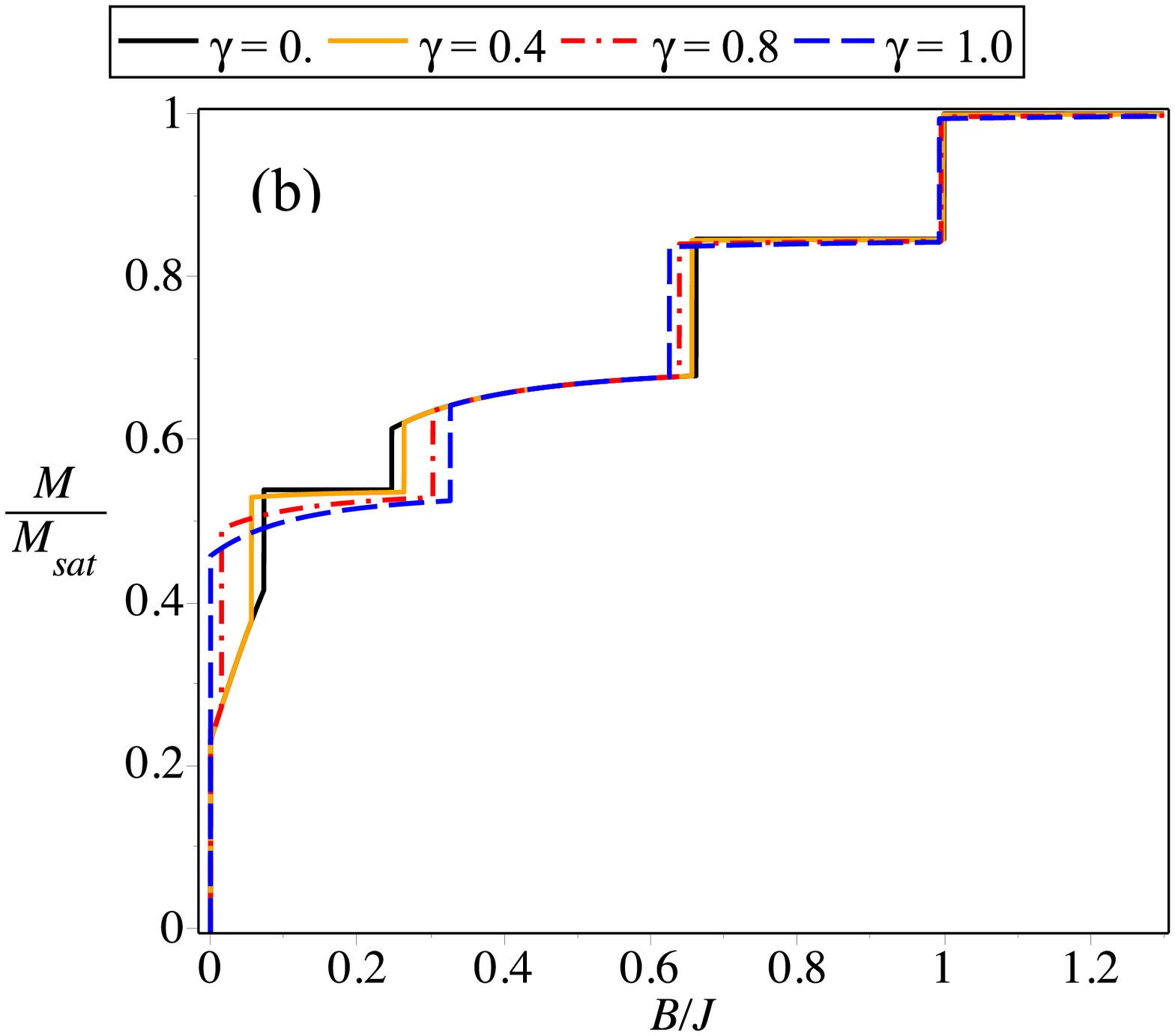}
\protect\caption{\label{fig10} Zero temperature normalized magnetization as a function
of magnetic field. Assuming fixed values: $J=1, g_{1}=8, g_{2}=2, g_{3}=2, g_{4}=4, K=1$
(a) For a range values of $\Delta$ and fixed $\gamma=0.5$. (b) For
a range values of $\gamma$ and fixed $\Delta=1.2$. $M_{sat}=\frac{1}{2}\left(g_{1}+g_{2}+\frac{1}{2}\left(g_{3}+g_{4}\right)\right)=13/2$.}
\end{figure}

\begin{figure}
\includegraphics[scale=0.3]{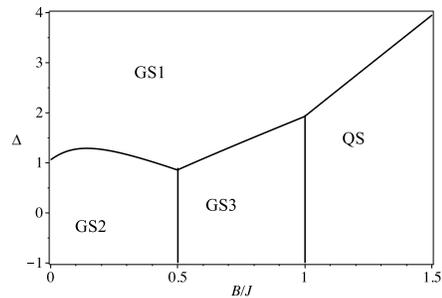}\caption{\label{fig:Phase-diagram}Zero temperature groud-states phase diagram $\Delta$ versus $B/J$ of the XYZ-Ising diamond-chain withe different g-factors for  $K=1$, $\gamma=0.5$ and g-factors $g_{1}=8$,
$g_{2}=2$, $g_{3}=2$ and $g_{4}=4$.}
\end{figure}

 We are not going to present the comprehensive analysis of all possible ground states and all types of magnetization curves for the XYZ-Ising diamond-chain with different g-factors, but let us just illustrate the ground state structure of the $T=0$ magnetization curve from Fig. (\ref{fig10}a) To be specific let us consider the magnetization curve for $J=1, K=1, g_1=8, g_2=2, g_3=2, g_4=4, \gamma=0.5$ and $\Delta=1.2$. The comparison of the different combinations of the spin-dimer eigenstates (Eq. (\ref{dim_es}) with $D_z=0$) with the orientation od the intermediate Ising spins gives rise to series of possible eigenstates for the chain. However, only few of them are realized in the case we considered here. First of all, the $B=0$ ground state is macroscopically degenerate for our choice of parameters. In this ground state all vertical quantum dimers are in $|\Psi_2\rangle$ state form Eq. (\ref{dim_es}) and all intermediate Ising spins can freely point either up or down. Thus, we have here $2^N$ configuration with the same energy. Arbitrary but non-zero magnetic field lifts this degeneracy. The system passes through the following ground state with the increasing the magnetic field which correspond to the course of the magnetization curve we are analyzing here:
 \begin{eqnarray}\label{dc_trans}
 |GS1\rangle\rightarrow |GS2\rangle\rightarrow|GS1\rangle\rightarrow|GS3\rangle\rightarrow|QS\rangle,
 \end{eqnarray}
 where the eigenstates with the corresponding magnetization and energies are
 \begin{widetext}
 \begin{eqnarray}\label{dc_gs}
 &&|GS1\rangle=\prod_{j=1}^N |\uparrow\rangle_j\otimes|\Psi_2\rangle_j,\\
  &&\qquad\;\; \mathcal{M}_1=\frac{Bg_-^2}{2\sqrt{B^2g_-^2+J^2}}+\frac 14\left(g_3+g_4\right),\;\; E_1=-\frac{J\Delta}{4}-\frac 12\sqrt{B^2g_-^2+J^2}-\frac B4\left(g_3+g_4\right),\nonumber\\
 &&|GS2\rangle=\prod_{j=1}^N |\downarrow\rangle_j\otimes|\Psi_4^+\rangle_j, \nonumber\\
 &&\qquad\;\; \mathcal{M}_2=\frac{\left(Bg_++2K\right)^2}{2B\sqrt{\left(Bg_++2K\right)^2+J^2\gamma^2}}-\frac 14\left(g_3+g_4\right),\;\; E_2=\frac{J\Delta}{4}-\frac 12\sqrt{\left(Bg_++2K\right)^2+J^2\gamma^2}+\frac B4\left(g_3+g_4\right),\nonumber \\
 &&|GS3\rangle=\prod_{j=1}^{N/2} |\downarrow\rangle_{2j-1}\otimes|\Psi_4\rangle_{2j-1}\otimes|\uparrow\rangle_{2j}\otimes|\Psi_4\rangle_{2j}, \nonumber\\
  &&\qquad\;\; \mathcal{M}_3=\frac{Bg_+^2}{2\sqrt{B^2g_+^2+J^2\gamma^2}}-\frac 14\left(g_3-g_4\right),\;\; E_3=\frac{J\Delta}{4}-\frac 12\sqrt{B^2g_+^2+J^2\gamma^2}+\frac B4\left(g_3-g_4\right),\nonumber\\
  &&|QS\rangle=\prod_{j=1}^N |\uparrow\rangle_j\otimes|\Psi_4^-\rangle_j, \nonumber\\
  &&\qquad\;\; \mathcal{M}_{QS}=\frac{\left(Bg_+-2K\right)^2}{2B\sqrt{\left(Bg_+-2K\right)^2+J^2\gamma^2}}+\frac 14\left(g_3+g_4\right),\;\; E_{QS}=\frac{J\Delta}{4}-\frac 12\sqrt{\left(Bg_+-2K\right)^2+J^2\gamma^2}-\frac B4\left(g_3+g_4\right),\nonumber
 \end{eqnarray}
 \end{widetext}
  here, $|\Psi_{2,4}\rangle_j$ stand for the corresponding eigenvectors of the spin-dimer (Eq. (\ref{dim_es}) with $D_z=0$) at j-th block and the $|\Psi_{4}^{\pm}\rangle_j$ is structurally the same $|\Psi_{4}\rangle_j$ eigenstate but taking into account the influence of the interaction with the neighboring Ising spins $\sigma_j$ and $\sigma_{j+1}$, which leads only to a modification of the coefficient $B_-$:
  \begin{widetext}
  \begin{eqnarray}
  &&|\Psi_4^{\pm}\rangle=\frac{1}{\sqrt{1+(B_-^{\pm})^2}}\left(|\uparrow\uparrow\rangle+B_-^{\pm}|\downarrow\downarrow\rangle\right),\\
  &&B_-^{\pm}
  =\frac{Bg_+\pm 2K-\sqrt{\left(Bg_+\pm2K\right)^2+J^2\gamma^2}}{J\gamma},\nonumber
  \end{eqnarray}
  \end{widetext}
  the arrows indicate the orientation of the corresponding Ising spins. Here several comments are in order. First of all, one can see the phenomenon of reentrant phase transition when the system with increasing the magnitude of the magnetic field enters the same ground state, $|GS1\rangle$, twice. It also clearly can be seen from the ground state phase diagram which is presented in Fig. (\ref{fig:Phase-diagram}) . The return to the ground state  $|GS1\rangle$ is possible due to non-linear magnetic field dependence of the corresponding magnetization and the energies of all ground states. One can also see from the Eq. (\ref{dc_gs})  that despite the visible ideal horizontal character of some part of the magnetization curve, they are not the magnetization plateaus, but just the very slowly growing part of the curve. Thus, the magnetization always has an explicit dependence on the magnetic field, unless $\gamma=0$ or/and $g_1=g_2$. The quasi-saturated state, $|QS\rangle$, at finite values of $\gamma$ has the magnetization asymptotically converging to $M_{sat}$. However, this value is inaccessible for finite values of magnetic field. Although most of quasi-plateaus appear below the saturation (last critical) field, there also exists
an alternative mechanism for quasi-plateau formation. Namely, the last plateau emerging above
the last critical field (which is in fact not a true saturation field) may change to the quasi-plateau due
to the XY anisotropy and consequently, the magnetization varies continuously above the last critical field
and it never reaches full saturation except for asymptotically infinite magnetic field.
\section{Conclusion}

In the present work, we have investigated in detail an absence of
actual plateaus in zero-temperature magnetization curves of quantum
spin clusters and chains. It has been convincingly evidenced that
strict plateaus may disappear from a magnetization process on assumption
that constituent spins of quantum spin clusters have different Landé
g-factors. More specifically, we have demonstrated this intriguing
feature on a few paradigmatic examples of quantum spin clusters such
as the spin-1/2 Heisenberg dimer, the mixed spin-(1/2,1) Heisenberg
dimer and the spin-1/2 Heisenberg trimer, whereas the same phenomenon
has been also found in the spin-1/2 Ising-Heisenberg diamond chain.
From this perspective, the absence of actual magnetization plateaus
due to the difference in Landé g-factors can be regarded as a general
feature of low-dimensional quantum antiferromagnets, because it emerges
whenever the total magnetization does not represent a conserved quantity
with well defined quantum spin numbers (the total magnetization need
not be proportional to the total spin). Accordingly, a smooth variation
of total magnetization can be attributed to a nonlinear dependence
of a few (or all) discrete energy levels on a magnetic field.

A few remarks are in order here, which might be useful for possible
experimental testing of this interesting phenomenon. Although the
magnetization curves of quantum spin clusters without strict plateaus
may mimic to a certain extent the magnetization curves of quantum
spin chains with continuous energy bands, it is obvious that the absence
of magnetization plateaus does not in turn mean a gapless excitation
spectrum. On the contrary, small magnetic spin clusters should always
have an energy gap, which can be easily experimentally tested by various
resonance techniques.

It is also worth noting that the deviation of magnetization from a strict plateau is proportional to a difference
between the Landé g-factors, which makes an experimental verification of this phenomenon more difficult. As a matter of fact, the most
of transition-metal ions as for instance Cu$^{2+}$, Ni$^{2+}$, Mn$^{2+}$, Cr$^{3+}$ or high-spin Fe$^{3+}$ with zero or totally quenched angular momentum
can be described by the notion of more or less isotropic quantum Heisenberg spins and hence, these magnetic ions usually have g-factors quite close
to the free electron value $g\approx2$  \cite{jomie,carlin,kahn}. Under these circumstances, it is customary to combine the local Landé g-factors of individual
magnetic ions into a single molecular g-factor as long as the isotropic Heisenberg exchange significantly prevails over the zero-field splitting, asymmetric
and/or antisymmetric exchange \cite{boca}. Many experimental studies focused on a magnetism of such oligonuclear complexes therefore simply ignore
different Landé g-factors of individual magnetic ions as the isotropic exchange is by far the most dominant coupling. However, a few transition-metal ions
with unquenched angular momentum may have much higher Land\'e g-factors due to a relatively strong spin-orbit coupling like for example the low-spin Fe$^{3+}$ ion
with typical value of $g \approx 2.8$ or Co$^{2+}$ ion with $g \approx 6.0$.\cite{jomie,carlin,kahn} Another possibility how to increase the difference of the Land\'e g-factors in oligonuclear complexes is to combine the almost isotropic transition-metal ion with highly anisotropic rare-earth ions, which may even possess much greater Land\'e g-factors (e.g. Dy$^{3+}$ typically
has $g \approx 20$) though this extraordinary large g-value usually correlates with a rather strong anisotropy in the exchange interaction.\cite{jomie,chibotaru,mironov}
Under the extreme situation, the XY-part of exchange coupling might be even of opposite sign than the Z-part (ferromagnetic versus antiferromagnetic)
as it has been recently reported for the heterodinuclear Cr$^{3+}$-Yb$^{3+}$ complex.\cite{mironov}

Last but not least, let us briefly comment on experimental implications
for the quantum spin clusters studied in the present work. The spin-1/2
Heisenberg dimer has previously proved its usefulness as the plausible
model of many homodinuclear Cu$^{2+}$-Cu$^{2+}$ complexes \cite{jomie,carlin,kahn,cucu}.
However, the difference between the local Landé g-factors in the homodinuclear
coordination compounds may only stem from a different coordination
sphere of individual magnetic ions, whereas this difference does not
exceed in most cases a few percent that would be insufficient for
an experimental testing. On the contrary, the considerable difference
in the local g-factors could be expected in heterobimetallic coordination
compounds, which are composed of the nearly isotropic magnetic ion
(e.g. Cu$^{2+}$, Ni$^{2+}$, high-spin Fe$^{3+}$, etc.) and the
highly anisotropic magnetic ion (e.g. Co$^{2+}$, low-spin Fe$^{3+}$,
etc.). Hence, the heterodinuclear Co$^{2+}$-Cu$^{2+}$ and Fe$^{3+}$-Cu$^{2+}$
complexes could be regarded as sought experimental realizations of the generalized
spin-1/2 Heisenberg dimer, which may display a substantial deviation
of the magnetization from a strict plateau as the g-factors of individual
magnetic ions may  even possess opposite signs due to a spin-orbit coupling
(e.g. $g_{{\rm Fe}}\approx-1.7$, $g_{{\rm Cu}}\approx2.1$ was reported in Ref. [\onlinecite{fecu}],
and negative g-factors of Co$^{2+}$ and Cu$^{2+}$ magnetic ions were investigated in Ref. [\onlinecite{ungur}]).
The similar conjecture can be formulated for experimental representatives of the mixed spin-(1/2,1)
Heisenberg dimer. As usual, the magnetic anisotropy in heterodinuclear
Cu$^{2+}$-Ni$^{2+}$ complexes does not cause a significant deviation
of the magnetization from a strict plateau \cite{nicu,cuni}, but
a rather large deviation could be expected instead in heterodinuclear
Co$^{2+}$-Ni$^{2+}$ coordination compounds with typical values of
the g-factors $g_{{\rm Co}}\approx5.9$ and $g_{{\rm Ni}}\approx2.3$
\cite{coni}. It can be also anticipated that the homotrinuclear Cu$^{2+}$-Cu$^{2+}$-Cu$^{2+}$
complexes \cite{cu3a,cu3b,cu3c} as experimental representatives of
the spin-1/2 Heisenberg trimer should not have a significant deviation
of the magnetization from a strict plateau unlike the heterotrinuclear
Cu$^{2+}$-Co$^{2+}$-Cu$^{2+}$ complexes \cite{cuco}.

\section{Acknowledgements}

V. O. and J. S. express their gratitude to the LNF-INFN for warm hospitality
during the work on the project. V. O. also acknowledges the partial
financial support form the grant by the State Committee of Science
of Armenia No. 13-1F343 and from the ICTP Network NET68. J. S. acknowledges the financial support
from the Scientific Grant Agency of Ministry of Education of Slovak Republic under contract
Nos. VEGA 1/0234/12 and VEGA 1/0331/15. O. R. thanks the Brazilian agencies FAPEMIG and CNPq for partial financial support.

\end{document}